\documentclass{amsart}%
\usepackage{amsmath}
\usepackage{amssymb}
\usepackage{amsthm}
\usepackage{amscd}
\usepackage[dvips]{graphicx}
\usepackage[mathscr]{eucal}
\newtheorem{Thm}{Theorem}[section]
\theoremstyle{definition}
\newtheorem{Theorem}[Thm]{Theorem}
\newtheorem{Lemma}[Thm]{Lemma}
\newtheorem{Corollary}[Thm]{Corollary}
\newtheorem{Proposition}[Thm]{Proposition}
\newtheorem{Definition}[Thm]{Definition}
\newtheorem{Example}[Thm]{Example}
\theoremstyle{remark}
\newtheorem{Remark}{Remark}

\font\ym=msbm10  
\newcommand{\Aut}{{\rm Aut}}

\newcommand{\A}{\text{\ym A}}
\newcommand{\B}{\text{\ym B}}

\newcommand{\R}{\text{\ym R}}
\newcommand{\bS}{\text{\ym S}}
\newcommand{\T}{\text{\ym T}}
\newcommand{\C}{\text{\ym C}}

\newcommand{\sB}{\mathscr B}

\newcommand{\sH}{\mathscr H}
\newcommand{\sL}{\mathscr L}
\newcommand{\sS}{\mathscr S}

\title[]
{Geometry of Quasi-Free States of CCR Algebras}
\author[Yamagami Shigeru]{Yamagami Shigeru}

\begin{document}
\maketitle   
\begin{center}
Division of Mathematics and Informatics
\end{center}
\begin{center}
Ibaraki University 
\end{center}
\begin{center} 
Mito, 310-8512, JAPAN 
\end{center}    
\begin{center}
\email{yamagami@mx.ibaraki.ac.jp}
\end{center}

\begin{abstract}
Geometric positions of square roots of quasi-free states 
of CCR algebras are investigated together with 
an explicit formula for transition amplitudes among them. 
\end{abstract}

\section*{Introduction}
In the operator algebraic approach to quantum physics, 
fundamental are the algebras associated to canonical 
(anti)commutation relations, which are known to be 
CCR or CAR algebras respectively. 
These are interesting from mathematical view points as 
well and have their own origins inside mathematics  
in connection with symplectic geometry or 
Clifford algebras. 

These quantum algebras are tied up with real experiments 
through probabilistic predictions, which can be described 
as transition probability between quantum states. 
In terms of quantum algebras, it is a standard recipe 
to recognize quantum states as normalized positive 
linear functionals. For commutative algebras, 
the notion of state in this sense is more or less 
equivalent to that of probability measure. 

Most naive quantum states are given by (normalized) vectors 
on which quantum algebras operate linearly and 
the transition probability between such states is 
postulated to be the squared modulus of their inner 
product. 

The extension of the notion of transition probability 
to general states are considered by several researchers, 
among them, let us notice the one introduced by 
A.~Uhlmann (\cite{U,AU}), 
which goes back to Kakutani-Bures analysis 
of measure-theoretical equivalence of product states. 
Viewing the observation in \cite{R}, 
Uhlmann's definition of transition probability 
takes the following form in the setting of W*-algebras: 
Let $\varphi$ and $\psi$ be normal states on a W*-algebra $M$ 
with $\varphi^{1/2}$ and $\psi^{1/2}$ the associated 
representing vectors in the positive cone of the standard 
Hilbert space $L^2(M)$. Then their (non-commutative) 
product $\varphi^{1/2}\psi^{1/2}$ is well-defined as 
a normal linear functional of $M$. If we denote by 
$|\varphi^{1/2}\psi^{1/2}|$ the positive part of 
$\varphi^{1/2}\psi^{1/2}$ in the polar decomposition, 
then the probability is equall 
to the value at the unit element $1 \in M$ 
of the positive linear functional 
$|\varphi^{1/2}\psi^{1/2}|$. 
Instead of taking positive parts, if simply evaluated 
the (not necessarily positive) form $\varphi^{1/2}\psi^{1/2}$, 
then we obtain another candidate for transition probability 
between states. 
Clearly this new transition probability 
(we call it transition amplitude simply because of 
its appearance) is not greater than 
the Uhlmann's one but both of them are reduced to 
the physically well-established one when 
restricted to vector states of full operator algebras. 

Returning to CCR or CAR algebras, among states, most studied 
are so-called quasifree states, which are free in the sense 
that they are determined in a canonical fashion 
and parametrized by their covariance operators 
(or covariance forms). 
Let us write $\varphi_S$ to stand for the quasifree state of 
a covariance operator $S$. 

In the case of CAR algebras, 
we have the following formula due to H.~Araki 
(\cite{Ar2}) for our transition amplitude. 
\[
(\varphi_S^{1/2}|\varphi_T^{1/2}) 
= \left( 
\det(1 - (P-Q)^2)
\right)^{1/8},  
\]
where the dilated projections $P$ and $Q$ are defined by 
\[
P = 
\begin{pmatrix}
S & \sqrt{S(1-S)}\\
\sqrt{S(1-S)} & 1-S
\end{pmatrix}, 
\qquad 
Q = 
\begin{pmatrix}
T & \sqrt{T(1-T)}\\
\sqrt{T(1-T)} & 1-T
\end{pmatrix}. 
\]

An analogous expression is possible to write down 
for CCR algebras as well under the condition that 
the relevant symplectic form is non-degenerate. 

Our main result in this paper is to establish 
a similar but undilated formula 
(see Theorem~\ref{finite} and Theorem~\ref{infinite} below) 
in a quite general form, 
which reveals a peculiar feature of quasifree states:  
To each covariance form $S$, we can associate 
a gaussian measure $\mu_S$ in such a way that 
$(\varphi_S^{1/2}|\varphi_T^{1/2})$ is equal to 
the Hellinger integral between $\mu_S$ and $\mu_T$. 
In other words, geometric positions of vectors 
$\varphi_S^{1/2}$ are exactly those of 
completely classical objects $\mu_S^{1/2}$.

\section{CCR Algebras}

Let $V$ be a real vector space and
$\sigma: V\times V \to \R$ be an alternating form.
The couple $(V,\sigma)$ is called 
a \textbf{presymplectic vector space}. 
When $\sigma$ is non-degenerate, it is called 
a \textbf{symplectic vector space}. 

Given presymplectic vector spaces $(V,\sigma)$ and 
$(V',\sigma')$, a linear map $\phi: V \to V'$ is 
said to be \textbf{presymplectic} if 
$\sigma'(\phi x,\phi y) = \sigma(x,y)$ for $x, y \in V$. 
When $(V',\sigma') = (V,\sigma)$ and $\phi$ is an isomorphism, 
it is called a presymplectic automorphism of $(V,\sigma)$. 
The group of presymplectic automorphisms of  
$(V,\sigma)$ is denoted by $\Aut(V,\sigma)$. 
When $(V,\sigma)$ is symplectic, 
all these maps are said to be symplectic. 

It is often useful to work with the complexification 
$V^\C$, which is furnished with the real structure 
$(x+iy)^* = x - iy$ so that $V$ is 
recovered as the real part of $V^\C$. 
Given a sesquilinear form $S$ on $V^\C$, we set 
$\overline S(z,w) = \overline{S(z^*,w^*)}$. Likewise, 
given a $\C$-linear transformation $\phi: V^\C \to V^\C$, 
we set ${\overline \phi}x = (\phi x^*)^*$. 

If the presymplectic form $\sigma$ is bilinearly extended to 
$V^\C$, then 
$h: V^\C \times V^\C \ni (z,w) \mapsto i \sigma(z^*,w)$ 
defines a hermitian form satisfying $\overline h = -h$. 
Conversely, a hermitian form $h$ on $V^\C$ satisfying 
${\overline h} = -h$ comes from
a presymplectic form $\sigma(x,y) = -ih(x,y)$ ($x, y \in V$). 

Associated to a presymplectic vector space $(V,\sigma)$, 
we introduce 
the \textbf{Weyl form of CCR algebra} 
as a unital *-algebra $C(V,\sigma)$ universally generated by 
the symbols $\{ e^{ix}; x \in V \}$ subject to the relations 
\[
(e^{ix})^* = e^{-ix}, 
\quad 
e^{ix} e^{iy} = e^{-i\sigma(x,y)/2} e^{i(x+y)}, 
\quad 
x, y \in V,
\]
which are the exponentiated form of 
the canonical commutation relations.
Note that $e^{i0}$ (the zero in the exponential represents 
the zero vector in $V$) is the unit element in the algebra. 

Since the Weyl form of CCR algebra is generated by unitaries 
$\{ e^{ix}\}$, any *-representation is automatically bounded. 
The operator-norm completion with repsect to 
all *-representations 
is then a C*-algebra $C^*(V,\sigma)$, which is referred to as 
the \textbf{CCR C*-algebra}. From the very definition, there is 
a one-to-one correspondance between *-representations 
of the Weyl form of CCR algebra on a Hilbert space $\sH$ and 
*-representations of $C^*(V,\sigma)$ on $\sH$. 
There is also a one-to-one correspondance between 
states on $C^*(V,\sigma)$ and 
states on the Weyl form of CCR algebra. 

Owing to the existence of Schr\"odinger type representations, 
we know that the family $\{ e^{ix} \}_{x \in V}$ is 
linearly independent in $C^*(V,\sigma)$. 

If we are given a presymplectic map 
$\phi: V \to V'$, it induces 
a *-homomorphism $C^*(V,\sigma) \to C^*(V',\sigma')$ 
by universality.  
In particular, $\Aut(V,\sigma)$ is imbedded into 
$\Aut(C^*(V,\sigma))$. 

A *-representation $\pi: C^*(V,\sigma) \to \sL(\sH)$ of 
a CCR C*-algebra on a Hilbert space $\sH$ is said to be 
\textbf{finite-dimensionally continuous} or \textbf{regular} 
if for any $\xi, \eta \in \sH$ 
and for any finite-dimensional subspace $W \subset V$, 
$(\xi|\pi(e^{ix})\eta)$ is a continuous function of 
$x \in W$. 

\section{Quasifree States}
\subsection{Polarizations}

\begin{Definition}
Let $(V,\sigma)$ be a presymplectic vector space. 
A positive form $S$ on $V^\C$ is called a \textbf{polarization} 
of $(V,\sigma)$ if it satisfies 
\[
S(x,y) - {\overline S}(x,y) = i \sigma(x^*,y)
\quad \text{for $x, y \in V^\C$.}
\]
Let $\text{Pol}(V,\sigma)$ be 
the set of polarizations of $(V,\sigma)$, 
which is a convex set with an obvious action of 
$\Aut(V,\sigma)$. 
\end{Definition}

\begin{Example}
For $V = \R^2$, possible presymplectic forms are
(redundantly) parametrized up to choice of bases by the matrix 
\[
\begin{pmatrix}
0 & 2\mu\\
-2\mu & 0  
\end{pmatrix},
\quad 
\mu \in \R.
\]
Then $S \in \text{Pol}(V,\sigma)$ 
is described by a matrix of the form 
\[
\begin{pmatrix}
z + x & y + i\mu\\ y - i\mu & z-x
\end{pmatrix}, 
\quad x^2 + y^2 + \mu^2 \leq z^2,  z\geq 0, 
\]
whence $\text{Pol}(V,\sigma)$ is identified with the region bounded 
by a half of a two-sheeted hyperboloid ($\mu \not= 0$) or 
a cone ($\mu=0$). 
Since 
\[
\Aut(V,\sigma) = 
\begin{cases}
\text{GL}(2,\R) &\text{if $\mu=0$,}\\
\text{SL}(2,\R) &\text{otherwise,}
\end{cases}
\]
orbits in $\text{Pol}(V,\sigma)$ constitute two or three parts 
according to $\mu \not= 0$ or $\mu=0$. 
\end{Example}

\begin{Lemma}[{\cite[Lemma~3.3]{AS}}]
Given a polarization $S$ of $(V,\sigma)$, set 
$(x,y)_S = S(x,y) + S(y^*,x^*)$ for $x, y \in V^\C$. 
Then $(\ ,\ )_S$ is a positive form on $V^\C$ satisfying 
(i) $(y^*,x^*)_S = (x,y)_S$ and (ii)
$|\sigma(x^*,y)|^2 \leq (x,x)_S\, (y,y)_S$ for $x, y \in V^\C$. 

Conversely, any positive form fulfilling these conditions 
comes from a polarization.  
\end{Lemma}

\begin{Corollary}
Let $V_S$ be the real Hilbert space associated to 
the positive form $S + \overline S$, i.e., 
$V_S$ is the completion of $V/(V \cap \ker(S+\overline S))$ 
with respect to the induced inner product. 
Then $\sigma$ is continuously extended to 
an alternating form $\sigma_S$ of $V_S$ so that 
the natural map $V \to V_S$ is presymplectic. 
\end{Corollary}

\subsection{Quasifree States}

\begin{Lemma}[Hadamard-Schur product]
Let $(a_{jk})$ and $(b_{jk})$ be positive 
semidefinite matrices of size $n$. Then the matrix 
with entries of component-wise multiplication 
$(a_{jk}b_{jk})_{1 \leq j, k \leq n}$ is 
positive semidefinite. 
\end{Lemma}

\begin{proof}
Express positive matrices 
as convex combinations of positive matrices of rank one. 
\end{proof}

\begin{Corollary}
For a positive semidefinite matrix $(a_{jk})$, 
the matrix $(e^{a_{jk}})$ is also positive semidefinite. 
\end{Corollary}

Given a polarization $S$, the following formula 
defines a state (called a \textbf{quasifree state}) 
on the CCR C*-algebra $C^*(V,\sigma)$. 
\[
\varphi_S(e^{ix}) = e^{-S(x,x)/2}, 
\quad x \in V.
\]
Since $\{ e^{ix} \}_{x \in V}$ are linearly independent in 
$C^*(V,\sigma)$, the linear functional $\varphi_S$ is 
well-defined on a dense subalgebra. 
To check the  positivity of $\varphi_S$, 
let $\{ x_j \}_{1 \leq j \leq n}$ be 
a finite family of vectors in $V$ 
and let $z_j \in \C$ ($1 \leq j \leq n$). 
Then  
\begin{align*}
\varphi_S\Bigl(
\left(
\sum_j z_j e^{ix_j}
\right)^*
\left(
\sum_k z_k e^{ix_k}
\right)
\Bigr)
&= \sum_{j,k} \overline{z_j} z_k 
e^{
-S(x_j-x_k,x_j-x_k)/2 
+ i\sigma(x_j,x_k)/2}\\
&= \sum_{j,k} 
\overline{e^{-S(x_j,x_j)/2}z_j}
e^{-S(x_k,x_k)/2}z_k
e^{S(x_j,x_k)}
\end{align*} 
is non-negative by the above corollary. 

Given a quasifree state $\varphi_S$, let $\pi_S$ be 
the associated GNS-representation. 
Then $\pi_S(e^{ix})$ 
is strongly continuous in $x \in V$ with respect to 
the topology induced from the inner product $(\ ,\ )_S$ 
because both of $S$ and $\sigma$ are continuous 
relative to $(\ ,\ )_S$. 

Given a C*-algebra $A$, let $L^2(A)$ be the standard 
Hilbert space of the W*-algebra $A^{**}$. We identify 
each $\varphi \in A^*_+$ with a normal functional of 
$A^{**}$ and use the notation $\varphi^{1/2}$ to stand for 
the associated representing vector 
in the positive cone of $L^2(A^{**})$. 

\begin{Proposition}[\cite{GMS}] 
Let $\phi: (V,\sigma) \to (V',\sigma')$ be 
a presymplectic map with the induced *-homomorphism 
$\Phi: C^*(V,\sigma) \to C^*(V',\sigma')$. 
Let $S, T \in \text{Pol}(V',\sigma')$ and assume that 
$\phi(V)$ is dense in $V'$ with respect to both of 
$S+\overline S$ and $T+\overline T$. 
Then 
\[
((\varphi_S\circ \Phi)^{1/2}| 
(\varphi_T\circ \Phi)^{1/2}) 
= (\varphi_{S\circ \phi}^{1/2}| 
\varphi_{T\circ \phi}^{1/2}).
\]
\end{Proposition}

If $\phi \in \Aut(V,\sigma)$, 
then $\Phi \in \Aut(C^*(V,\sigma))$ and 
\[
\varphi_S^{1/2} \mapsto \varphi_{S\circ \phi}^{1/2}
\]
for various $S \in \text{Pol}(V,\sigma)$ gives rise to 
a unitary operator on the subspace of $L^2(C^*(V,\sigma))$ 
spanned by 
$\{ \varphi^{1/2}_S; S \in \text{Pol}(V,\sigma) \}$.  

Let $S$ be a polarization of $(V,\sigma)$ 
and $\bS$ be the positive 
operator on $V_S^\C$ defined by $S(x,y) = (x, \bS  y)_S$, 
where $S$ is identified with the induced polarization of 
$(V_S,\sigma_S)$. 
We say that $S$ is \textbf{in the boundary} unless 
$\ker\bS = \{ 0\} = \ker(1-\bS)$. 
By the theory of Fock representations and 
the method of doubling or purification 
(cf.~\S 5 below), we know that 
$\varphi_S$ is a pure state if and only if 
$S$ is extremal in $\text{Pol}(V,\sigma)$, 
i.e., $\bS$ is a projection. 

If $S$ is not in the boundary, 
we can define a one-parameter group 
of unitaries $\{ e^{itH} \}$ on $V_S^\C$ by 
$e^{itH} = \bS^{-it}(1-\bS)^{it}$. Note that the infinitesimal 
generator $H$ (which is a self-adjoint operator) 
satisfies $\overline H = - H$. 
In accordance with the functional calculus notation for 
positive forms, the operator $e^{itH}$ is also denoted by 
$S^{-it} {\overline S}^{it}$. 

\begin{Proposition}[{\cite[\S 3]{Ar}}]\label{KMS}
Assume that $S$ is not in the boundary. 
Then 
\begin{enumerate}
\item 
The family 
$\{ e^{itH} = S^{-it}{\overline S}^{it} \}_{t \in \R}$ 
is, when restricted to $V_S$, 
a one-parameter group of presymplectic transformations 
of $(V_S,\sigma_S)$ such that 
$S(e^{itH}x,e^{itH}y) = S(x,y)$ for $x, y \in V_S^\C$. 
\item 
The associated automorphism group of $C^*(V_S,\sigma_S)$ 
satisfies the KMS-condition 
for the quasifree state $\varphi_S$: 
\[
\varphi_S^{it} e^{ix} \varphi_S^{-it} = \exp(ie^{itH}x) 
\]
for $x \in V_S$ and $t \in \R$. 
(See \cite{AA}
for the meaning of notation $\varphi_S^{it}$.) 
\end{enumerate}
\end{Proposition}

\subsection{Central Decoposition of Quasifree States}

We assume that $V$ is endowed with 
a topology of Hilbert space 
($V$ is then said to be hilbertian). 
A polarization $S$ of $(V,\sigma)$ is said to be 
\textbf{admissible} if $S+\overline S$ gives the topology 
of $V$. 
Let $V_0 = \ker \sigma$ be the central part of $V$ 
and 
$S_0$ be the restriction of $S$ to $V_0^\C$. 
Let $V_1 = V \ominus_S V_0$ be 
the $(S+\overline S)$-orthogonal complement of $V_0$. 

We choose an auxiliary linear map of Hilbert-Schmidt class
$\Theta: L \to V_0$ with $\ker\Theta = \{ 0\}$ and 
having a dense range
($L$ being a real Hilbert space). 
Then we can realize the state $\varphi_{S_0}$ by 
a gaussian measure $\nu_{S_0}$ on the topological dual 
space $\Omega = (\Theta L)^*$, where 
$\Theta L$ is furnished with the topology induced 
from the norm 
$\| v_0\|_\Theta = \| \Theta^{-1} v_0\|$: 
\[
\int_\Omega e^{i\omega(x_0)} \nu_{S_0}(d\omega) 
= e^{-S(x_0)/2},
\quad x_0 \in \Theta L.
\]
and 
\[
\overline{C^*(V_0)\varphi_{S_0}^{1/2}} 
= L^2(\Omega,\nu)
\quad
\text{by}
\quad 
e^{ix_0}\varphi_{S_0}^{1/2} 
\longleftrightarrow 
e^{i\omega(x_0)} \sqrt{\nu_{S_0}(d\omega)}. 
\]

\begin{Remark}
For an infinite-dimensional $V_0$, 
$\nu_{S_0}(V_0) = 0$, where 
$V_0$ is imbedded into the space $\Omega = (\Theta L)^*$ 
through the inner product $S_0$. 
\end{Remark}

We now introduce a C*-algebra $C^*(V/\omega,\sigma)$ 
for each $\omega \in \Omega$, 
which is the quotient of the CCR C*-algebra 
$C^*(V_1 + \Theta L,\sigma)$ by imposing the condition
$e^{ix_0} = e^{i\omega(x_0)} 1$ ($x_0 \in \Theta L$).
Namely, if we denote the quotient image of $e^{ix}$ by 
$e_\omega^{ix}$, the C*-algebra $C^*(V/\omega,\sigma)$ 
is generated by unitaries 
$\{ e_\omega^{ix}; x \in V_1 + \Theta L \}$ 
subject to the relations 
\[
e_\omega^{ix} e_\omega^{iy} 
= e^{-i\sigma(x,y)/2} e_\omega^{i(x+y)} , 
\quad
e_\omega^{ix_0} = e^{i\omega(x_0)} 1
\]
for $x, y \in V_1 + \Theta L$ and $x_0 \in \Theta L$. 
Notice here that $C^*(V/\omega,\sigma)$ depends not only on 
$\Theta$ but also on $S$ through the choice 
$V_1 + \Theta L \subset V$. 
Clearly 
\[
C^*(V/\omega,\sigma) 
\ni e^{ix_1} \mapsto 
e^{ix_1} \in C^*(V_1,\sigma_1), 
\quad x_1 \in V_1 
\]
gives an isomorphism of C*-algebras 
with $\sigma_1 = \sigma|_{V_1\times V_1}$ 
the induced symplectic form on $V_1$, whence 
we can define a state $\varphi_{S,\omega}$ of 
$C^*(V/\omega,\sigma)$ so that it corresponds to 
the quasifree state $\varphi_{S_1}$ of 
$C^*(V_1,\sigma_1)$ via the above isomorphism: 
\[
\varphi_{S,\omega}(e_\omega^{ix}) 
= e^{-S(x_1,x_1)/2} e^{i\omega(x_0)}
\quad 
\text{for $x = x_1 + x_0 \in V_1 + \Theta L$.}
\] 

\begin{Proposition}\label{central}
Given a polarization $S$ of the presymplectic vector space 
$(V,\sigma)$ with $\Theta: L \to V_0$ as above, 
we have the following unitary isomorphism 
\begin{align*}
\overline{
C^*(V,\sigma) \varphi_S^{1/2} C^*(V,\sigma)
} 
\quad&\cong\  
\int_\Omega^\oplus \nu_{S_0}(d\omega)\, 
\overline{
C^*(V/\omega,\sigma) 
\varphi_{S,\omega}^{1/2}
C^*(V/\omega,\sigma)
}\\
e^{ix} \varphi_S^{1/2} e^{iy} 
&\longleftrightarrow 
\int_\Omega^\oplus \nu_{S_0}(d\omega)\, 
e_\omega^{ix} \varphi_{S,\omega}^{1/2} 
e_\omega^{iy}
\qquad
(x, y \in V_1 + \Theta L).
\end{align*}
\end{Proposition}

\begin{proof}
Step 1. By splitting out the pure state part 
(corresponding to the spectral range $\{ 0, 1\}$) from $S$, 
we may assume that $S$ is not in the boundary
(the isomorphism in question turns out to be the identity 
on the factor arizing from pure state part). 

Step 2. If $y \in V_1 + \Theta L$ is an entirely analytic 
vector for the self-adjoint operator $H = \log((1-\bS)/\bS)$,  
\[
\varphi_{S,\omega}^{1/2} e_\omega^{iy} 
= e_\omega^{iS^{-1/2}{\overline S}^{1/2}y} 
\varphi_{S,\omega}^{1/2}
\quad 
\text{with}\ 
S^{-1/2} {\overline S}^{1/2} = e^{-H/2}.
\]
Note here that $e_\omega^{iz}\varphi_{S,\omega}^{1/2}$ 
is well-defined as an entirely analytic function of 
$z \in (V_1 + \Theta L)^\C$ (cf.~Lemma~\ref{fock} below). 

By the isomorphism 
$C^*(V_1,\sigma_1) \ni e^{ix_1} \mapsto e_\omega^{ix_1} 
\in C^*(V/\omega,\sigma)$, 
$\varphi_{S,\omega}$ corresponds to $\varphi_{S_1}$, 
whence 
\[
\varphi_{S,\omega}^{1/2} e_\omega^{iy_1} 
= e_\omega^{iS_1^{-1/2}{\overline{S_1}}^{1/2} y_1} 
\varphi_{S,\omega}^{1/2}
\]
for $y_1 \in V \ominus_S V_0$. Since 
$S^{-1/2}{\overline S}^{1/2}y 
= y_0 + S_1^{-1/2} {\overline{S_1}}^{1/2}y_1$ 
for $y = y_0 + y_1$ and $e^{iy_0}$ is in the center, 
the assertion follows. 

Step 3. For $x \in V_1 + \Theta L$, 
\begin{align*}
(\varphi_S^{1/2}|e^{ix}\varphi_S^{1/2})
&= e^{-S(x_0)/2} e^{-S(x_1)/2}
= \int_\Omega \nu_{S_0}(d\omega) e^{i\omega(x_0)} 
e^{-S(x_1)/2}\\
&= \int_\Omega \nu_{S_0}(d\omega) 
(\varphi_{S,\omega}^{1/2}| 
e_\omega^{ix} \varphi_{S,\omega}^{1/2}).
\end{align*}
\end{proof}

\begin{Corollary}
Let $T$ be another admissible polarization of $(V,\sigma)$ 
such that 
$(V \ominus_T V_0) + \Theta L = (V \ominus_S V_0) + \Theta L$ 
and $\nu_T$ is equivalent to $\nu_S$. 
Then 
\[
(\varphi_S^{1/2}|\varphi_T^{1/2}) 
= \int_\Omega \sqrt{\nu_{S_0}\nu_{T_0}}(d\omega) 
(\varphi_{S,\omega}^{1/2}|
\varphi_{T,\omega}^{1/2}).
\]
\end{Corollary}

\section{Finite-Dimensional Analysis}

Let $(V,\sigma)$ be a finite-dimensional 
presymplectic vector space, 
which is assumed throughout this section 
unless otherwise stated, and we shall fix 
a Lebesgue measure on $V$ once for all. 

\subsection{Hilbert Algebras}
Let $\sS(V)$ be the space of rapidly decreasing functions 
on $V$. For $f \in \sS(V)$, regard the integral  
\[
\int_V f(x) e^{ix}\,dx 
\]
as defining a virtual element in $C^*(V,\sigma)$ 
(more precisely, it belongs to the von Neumann algebra generated 
by a regular representation), 
which suggests a *-algebra structure in 
the vector space $\sS(V)$: 
\[
(f*g)(z) = \int_V f(z') g(z-z') e^{-i\sigma(z',z)/2}\,dz', 
\quad
f^*(x) = \overline{f(-x)}.
\] 

It is immediate to check that these operations 
in fact make $\sS(V)$ into a *-algebra 
(denoted by $\sS(V,\sigma)$), 
which turns out to be a Hilbert algebra with respect to 
the inner product 
\[
(f|g) = \int_V \overline{f(x)} g(x)\,dx.
\]
The (associated) trace functional is then defined by 
\[
\tau: \sS(V) \ni f \mapsto f(0) \in \C.
\]
Formally this is equivalent to requiring  
$\tau(e^{ix}) = \delta(x)$ for $x \in V$ 
($\delta(x)$ being the delta function 
with respect to the preassigned Lebesgue measure). 
Moreover, the multiplier product by $e^{ix} \in C^*(V,\sigma)$ 
leaves $\sS(V)$ invariant and we have 
the multiplier product formula 
\[
(e^{ix}fe^{iy})(z) = 
f(z-x-y) e^{i(\sigma(x,y) - \sigma(x,z) - \sigma(z,y))/2}
\]
for $x, y, z \in V$, 
which follows from the identity 
\[
e^{ix}
\left(
\int_V f(z) e^{iz}\,dz
\right) e^{iy} = \int_V 
f(z-x-y) e^{i(\sigma(x,y) - \sigma(x,z) - \sigma(z,y))/2} e^{iz}dz. 
\]
In other words, if we denote by $C^*_\text{reg}(V,\sigma)$ 
the C*-closure of $\sS(V,\sigma)$, then 
$C^*(V,\sigma)$ is identified with a multiplier subalgebra of 
$C^*_\text{reg}(V,\sigma)$. 

Since the quasifree state $\varphi_S$ gives rise to the functional 
\[
\int_V f(x) e^{ix}\,dx \mapsto 
\int_V f(x) e^{-S(x,x)/2}\,dx,
\]
the associated density operator $\rho_S$, i.e., 
$\rho_S \in \sS(V,\sigma)$ satisfying 
$\tau(\rho_S*f) = \varphi_S(\int dx f(x) e^{ix})$ for 
$f \in \sS(V)$, is given by the function 
$\rho_S(x) = e^{-S(x,x)/2}$ ($x \in V$). 

To get an expression for the square root of $\rho_S$ 
in $\sS(V,\sigma)$, 
we need therefore to seek for a function $f: V \to \C$ 
satisfying $f(-x) = \overline{f(x)}$ and 
\[
e^{-S(x,x)/2} = \int_V f(y) f(x-y) e^{i\sigma(x,y)/2}\,dy 
= (f*f)(x)
\]
for $x \in V$. 

\subsection{Non-Degenerate $\sigma$}
From here on $\sigma$ is assumed to be non-degenerate 
for the time being. 
Let $\bS$ be the operator representing the polarization $S$ 
with respect to the inner product $(\ ,\ )_S$. 
From the relation $\bS + \overline{\bS} = 1$, 
we read off the following spectral property on $\bS$: 
If $\xi$ is an eigenvector of eigenvalue $\lambda$, 
then so is $\xi^*$ 
with eigenvalue replaced by $1 - \lambda$, 
i.e., ${\bS}\xi^* = (1-\lambda)\xi^*$. 
To utilize this property, we assume $0 \leq \lambda < 1/2$,
normalize the eigenvector $\xi$ and introduce orthonormal vectors in $V$ by 
\[
e = \frac{\xi + \xi^*}{\sqrt{2}}, 
\quad 
f = \frac{\xi - \xi^*}{\sqrt{2} i}, 
\quad 
\xi = \frac{e+if}{\sqrt{2}}.
\]

Relative to the basis $\{ e, f\}$, $\bS$ is 
represented by the matrix 
\[
\bS = 
\begin{pmatrix}
1/2 & i\mu\\
-i\mu & 1/2 
\end{pmatrix}
\quad
\text{with}\quad 
\overline{\bS} = 
\begin{pmatrix}
1/2 & -i\mu\\
i\mu & 1/2 
\end{pmatrix}
\ 
\text{and}
\   
\bS - \overline{\bS} 
= i 
\begin{pmatrix}
0 & 2\mu\\
-2\mu & 0
\end{pmatrix}, 
\]
where $2\mu \equiv 1 - 2\lambda$. 

Consequently the canonical (Liouville) measure is of the form
$2\mu dsdt$ with respect to the (partial) coordinates 
$(s,t) \in \R^2$ representing the vector $se + tf$ 
in a two-dimensional subspace of $V$, whereas 
the reference measure is of the form $2mdsdt$ with $m>0$. 
As the relevant forms are calculated to be 
\[
S(se+tf, se+tf) = \frac{1}{2}(s^2 + t^2), 
\quad 
i \sigma(se+tf, s'e + t'f) = 2i\mu(st' - s't),
\]
the equation to determine $f$ takes the expression 
\[
e^{-(s^2+t^2)/4} 
= 2m \int_{\R^2} f(s',t') f(s-s', t-t') 
e^{i\mu(st' - s't)}\, ds'dt'
\]
with the hermiticity condition given by 
$f(-s,-t) = \overline{f(s,t)}$. 
We shall deal with a slightly more general situation: 
for $f, g \in \sS(V)$, consider 
\[
(f*g)(se+tf) = 2m \int_{\R^2} f(s',t') g(s-s',t-t') 
e^{i\mu(st'-s't)}\, ds'dt'. 
\]
If we write
\[
f(s,t) = \frac{1}{(2\pi)^2} 
\int_{\R^2}
\widehat f(\xi,\eta) e^{is\xi + it\eta}\,d\xi d\eta
\]
with the Fourier transform $\widehat f$ defined by 
\[
\widehat f(\xi,\eta) = \int_{\R^2}
f(s,t) e^{-is\xi - it\eta}\,dsdt
\]
and similarly for $g$, then 
\[
(f*g)(s,t) = 
\frac{2m}{(2\pi)^2} \int_{\R^2} 
{\widehat f}(\xi,\eta) 
{\widehat g}(\xi - \mu t, \eta + \mu s) 
e^{is\xi + it\eta}\, d\xi d\eta.
\]

For the choice 
\[
{\widehat f}(\xi,\eta) = A e^{-a(\xi^2+\eta^2)/2\mu}, 
\quad
{\widehat g}(\xi,\eta) = B e^{-b(\xi^2+\eta^2)/2\mu}
\]
with $A, B, a, b$ positive reals, explicit computations 
are worked out by gaussian integrals: The results are 
\[
f(s,t) = \frac{\mu A}{2\pi a} e^{-\mu(s^2+t^2)/2a}, 
\quad 
g(s,t) = \frac{\mu B}{2\pi b} e^{-\mu(s^2+t^2)/2b}
\]
and 
\[
(f*g)(s,t) = 
\frac{ABm\mu}{\pi(a+b)} 
e^{-\mu(s^2+t^2)/2(a*b)}, 
\]
where 
\[
a*b = \frac{a+b}{ab + 1}. 
\]
Notice here that the function 
$h(s,t) = \exp(-\mu(s^2+t^2)/2c)$ with $c>0$ 
defines a positive element 
\[
h = 2m \int_{\R^2} h(s,t) \pi(e^{i(se+tf)})\, dsdt
\]
in any regular representation $\pi$ of $C^*(V,\sigma)$ 
if and only if the associated positive functional 
$\tau(h\cdot)$ is a quasifree state, which means that  
\[
\frac{\mu}{c} (s^2 + t^2) 
= 
\begin{pmatrix}
s & t
\end{pmatrix}
\begin{pmatrix}
z & i\mu\\
-i\mu & z
\end{pmatrix}
\begin{pmatrix}
s\\
t
\end{pmatrix}
\]
with $z = \mu/c > 0$ satisfying $\mu^2 \leq z^2$, namely 
$c \leq 1$. 

In particular, for the choice $a = b = c > 0$ and $A = B = C > 0$, 
\[
(f*f)(se+tf) = \frac{C^2m\mu}{2\pi c} e^{-\mu(s^2+t^2)/2(c*c)},
\]
where $c*c = 2c/(c^2+1) \leq 1$ for any $c>0$ as being expected 
(the square of a hermitian element being positive). 
If this is identified with 
$e^{-(s^2+t^2)/4}$, we find a solution 
\[
f(s,t) = \sqrt{\frac{\mu}{2\pi cm}} e^{-\mu(s^2+t^2)/2c} 
\]
satisfying 
$0 < c \leq 1$ (recall $0 < 2\mu \leq 1$) by the choice 
\[
c = \frac{2\mu}{1 + \sqrt{1 - 4\mu^2}}, 
\quad 
C = \frac{\sqrt{2\pi c}}{\sqrt{m\mu}}.
\]

From the expression
\[
\bS +\overline{\bS} 
+ 2\sqrt{\bS \overline{\bS}} 
= 
\begin{pmatrix}
1 + \sqrt{1 - 4\mu^2} & 0\\
0 & 1 + \sqrt{1-4\mu^2}
\end{pmatrix},
\]
we see that 
\[
\frac{\mu}{2c}(s^2+t^2) = 
\frac{1}{4} 
\begin{pmatrix}
s & t
\end{pmatrix}
\left(\bS +\overline{\bS} 
+ 2\sqrt{\bS \overline{\bS}} 
\right)
\begin{pmatrix}
s\\
t
\end{pmatrix}.
\]

The last relation immediately gives rise to the general formula 
for $\rho_S^{1/2}$ 
($ = \int_V \rho_S^{1/2}(x) e^{ix}\,dx$): 
\[
\rho_S^{1/2}(x) 
= \frac{1}{\sqrt{N_S}}
\exp\left(
-\frac{1}{4} \biggl(S + \overline{S} + 2\sqrt{S\overline{S}}\biggr)(x,x) 
\right)
\]
with 
\[
N_S = \int_V 
\exp\left(
-\frac{1}{2}\biggl(S+\overline{S} + 2\sqrt{S\overline S}\biggr)(x,x)
\right)\,dx. 
\]
Here the geometric mean $\sqrt{S\overline S}$ of 
positive forms is in the sense of Pusz-Woronowicz 
(\cite{PW}).
 
The expression for the normalization constant is obtained 
from the identity 
\[
\int_V \rho_S^{1/2}(x) \rho_S^{1/2}(x)\,dx = \rho_S(0) = 1.
\]
Compare this with 
\[
\rho_S(x) = e^{-S(x,x)/2} = 
\exp\left(
-\frac{1}{4}(S+\overline{S})(x,x)
\right),
\]
which suggests the following rule how  
the exponential part is changed 
under the replacement $\rho_S \Longrightarrow \rho_S^{1/2}$: 
$S + \overline{S} \Longrightarrow 
\left(
\sqrt{S} + \sqrt{\overline{S}} 
\right)^2$. 

Also remark that 
$(S + \overline{S} + 2\sqrt{S\overline{S}})(x,x) 
= (x,x)_{S + \sqrt{S\overline{S}}}$ 
with $S + \sqrt{S\overline{S}} \in \text{Pol}(V,\sigma)$. 

\subsection{Non-Degenerate $S + \overline S$} 
From here on, the alternating form $\sigma$ 
is relaxed to be presymplectic 
and consider a polarization $S$ such that $(\ ,\ )_S$ 
is non-degenerate. The kernel of $\sigma$ is then captured 
as the eigenspace of $\bS$ corresponding to an eigenvalue $1/2$. 
To deal with the degenerate part of $\sigma$, let $\eta \in V$ be 
a normalized eigenvector (${\bS}\eta = \eta/2$ and $(\eta,\eta)_S = 1$) 
and assume that the reference measure 
is of the form $m_0dt$ 
with respect to the coordinate $t$ 
representing the vector $t\eta$ in $V$. 

Then the equation for $f$ takes the form 
\[
e^{-t^2/4} = m_0 \int_{-\infty}^\infty f(s) f(t-s)\,ds
\]
with the positive solution given by 
\[
\rho_S^{1/2}(t\eta) = f(t) = N_0^{-1/2} e^{-t^2/2} 
= N_0^{-1/2} e^{-S(t\eta,t\eta)}. 
\]
Here the normalization constant is given by 
\[
N_0 = m_0 \int_{-\infty}^\infty e^{-2S(t\eta,t\eta)} dt
= m_0\sqrt{\pi}. 
\]
Thus the formula for $\rho_S^{1/2}$ 
remains valid even for the degenerate 
case $\mu = 0$ as far as $S + \overline S$ is non-degenerate. 

For $S, T \in \text{Pol}(V,\sigma)$, let 
$A$ and $B$ be positive quadratic forms on $V$ defined by 
\begin{align*}
A(x) &= \frac{S(x,x)+\overline{S}(x,x)}{2} 
+ \sqrt{S\overline{S}}(x,x),\\ 
B(x) &= \frac{T(x,x)+\overline{T}(x,x)}{2} 
+ \sqrt{T\overline{T}}(x,x).
\end{align*}
Assume that both of $A$ and $B$ are non-degenerate on $V$. 
If the multiplier product formula is applied to 
\[
\rho_S^{1/2}(x) = 
N_S^{-1/2}
e^{-A(x)/2}, 
\quad 
N_S = \int_V e^{-A(x)}\,dx
\]
and to a similar expression for $\rho_T^{1/2}(x)$, 
we obtain  
\[
\tau\left( 
\rho_S^{1/2} e^{ix} \rho_T^{1/2}e^{iy} 
\right) 
= \frac{1}{\sqrt{N_S N_T}} 
\int_V e^{-A(z-x)/2 - i\sigma(z,x)/2} 
e^{-B(z+y))/2 + i\sigma(z,y)/2} dz 
\]
for $x, y \in V$. 
In particular, we have 
\[
(\varphi_S^{1/2}|\varphi_T^{1/2})
= \frac{\int_V e^{-(A(x)+B(x))/2}\,dx}
{\sqrt{\int_V e^{-A(x)}\,dx \int_V e^{-B(x)}\,dx}}. 
\]
Notice that the right hand side does not depend on 
the choice of volume form $dx$ 
in so far as it is translationally invariant. 
Also note that the right hand side is of the form 
$(\xi|\eta)$ with $\xi, \eta$ unit vectors in 
$L^2(V)$ (here $V$ being assumed to be finite-dimensional).

Thus the relative position of vectors 
$\{ \varphi_S^{1/2} \}_{S \in \text{Pol}'(V,\sigma)}$ 
in $L^2(C^*(V,\sigma))$ is same with that
of unit vectors $\{ \xi_S \}_{S \in \text{Pol}'(V,\sigma)}$ 
in $L^2(V)$, where 
\[
\xi_S = \frac{1}{\sqrt{\int_V e^{-A(x)}\,dx}} e^{-A(x)/2}\sqrt{dx} 
\]
and $\text{Pol}'(V,\sigma)$ denotes the set of polarizations 
$S$ such that $(\ ,\ )_S$ is non-degenerate. 

We now rewrite the transition amplitude formula 
into a form which can be used 
without the non-degeneracy restriction on polarizations. 
Choose an auxiliary inner product $(\ |\ )$ in $V$ so that 
$A$ and $B$ are represented by commuting positive invertible 
operators $\A$ and $\B$ respectively
($(\ ,\ )_S + (\ ,\ )_T$ can be used for example).
The reference measure is then normalized regarding 
the inner product $(\ |\ )$; use the ordinary Lebesgue measure 
associated to an orthonormal basis of $(\ |\ )$.  
Then 
\[
N_S = \int_V e^{-(x|{\A}x)}\,dx 
= \frac{\pi^{n/2}}{\sqrt{\det({\A})}}
\]
and we have 
\begin{align*}
(\varphi_S^{1/2}|\varphi_T^{1/2}) 
&= \frac{(\det{\A\B})^{1/4}}{\pi^{n/2}} 
\int_V e^{-(x|{\A}x)/2 - (x|{\B}x)/2}\ dx\\
&= \frac{\det^{1/4}(\A\B)}
{\sqrt{\det(\A/2 + \B/2)}}
= \sqrt{
\det\left(
\frac{2\sqrt{\A\B}}{{\A} + {\B}}
\right)
}.
\end{align*}
With the help of Pusz-Woronowicz' functional calculus, 
the last formula takes a coordinates-free (independent of 
the choice of auxiliary inner products) expression: 
\[
(\varphi_S^{1/2}|\varphi_T^{1/2}) = 
\sqrt{\det\left( 
\frac{2\sqrt{AB}}{A+B}
\right)}.
\]
See Appendix~A for the meaning of determinant in the right 
hand side. 

\begin{Example}
Let $\sigma$ be defined on $V = \R^2$ by the matrix 
$
\begin{pmatrix}
0 & 2\mu\\
-2\mu & 0
\end{pmatrix}$ with $\mu \in \R$ and consider a polarization 
$S$ in the boundary of $\text{Pol}(V,\sigma)$. 
From the spectral expression discussed above, we see that 
\[
\sqrt{S\overline{S}} = 
\begin{cases} 0 &\text{if $\mu \not= 0$,}\\
S &\text{if $\mu = 0$.}
\end{cases}
\]
 
Thus, if $S$ is associated to a matrix
$
\begin{pmatrix}
z + x & y + i\mu\\
y - i\mu & z-x
\end{pmatrix}$ ($z^2 = x^2 + y^2 + \mu^2$), 
then the quadratic form $A$ corresponds to the matrix 
$\begin{pmatrix}
z+x & y\\
y & z-x
\end{pmatrix}$ for $\mu \not= 0$, whereas it is multiplied by 
the factor $2$ for $\mu = 0$.

Notice that the matrix
$
\begin{pmatrix}
z + x & y + i\mu\\
y - i\mu & z-x
\end{pmatrix}$ ($z^2 = x^2 + y^2 + \mu^2$), 
does not commute with its complex conjugate 
unless $(x,y,\mu) = (0,0,0)$. 

Let $S'$ be another boundary polarization with the associated 
quadratic form denoted by $A'$. Then both of $A$ and $A'$ are 
positive definite for $\mu \not= 0$ and we have 
\[
(\varphi_S^{1/2}|\varphi_{S'}^{1/2}) 
= 2\frac{(z^2-x^2-y^2)^{1/4} ({z'}^2 - {x'}^2 - {y'}^2)^{1/4}}
{\sqrt{(z+z')^2 - (x+x')^2 - (y+y')^2}}. 
\]
\end{Example}

\subsection{Degenerate Case}
Now we shall remove the non-degeneracy condition 
on polarizations 
and extend  the transition amplitude formula 
between quasifree states to 
the general (but finite-dimensional) case. 
Let $(V,\sigma)$ be 
a finite-dimensional presymplectic vector space. 

\begin{Lemma}
Let $(V,\sigma)$ be a general presymplectic vector space. 
Let $S, T \in \text{Pol}(V,\sigma)$. If 
$\ker(S+\overline S) \not= \ker(T + \overline T)$, then 
$(\varphi_S^{1/2}|\varphi_T^{1/2}) = 0$.  
\end{Lemma}

\begin{proof}
We may assume that there is 
$x \in V$ such that $(x,x)_S = 0$ and $(x,x)_T \not= 0$. 
Consider the homomorphism $\pi: C^*(\R x,0) \to C^*(V,\sigma)$ induced from 
the presymplectic map $\R x \subset V$. Then it is easy to see that 
$\varphi_S\circ\pi$ is represented by the Dirac measure concentrated at $0 \in \R$ whereas 
$\varphi_T\circ\pi$ is described 
by a gaussian measure on $\R$. 
Thus by the transition amplitude inequality 
(\cite[Lemma~4.1]{GMS}) 
\[
0 \leq (\varphi_S^{1/2}| \varphi_T^{1/2}) \leq 
((\varphi_S \circ \pi)^{1/2}| (\varphi_T\circ \pi)^{1/2}) = 0.
\]
\end{proof}

As a result, 
we see that the transition amplitude formula is valid 
if $\ker(\ ,\ )_S \not= \ker (\ ,\ )_T$.
So assume that these kernels (denoted by $K$) coincide. 

Let $V'$ be the quotient space of $V$  
by $K$ with $\sigma'$, $S'$ and $T'$ the induced forms on 
$V'$. Let $\pi: C^*(V,\sigma) \to C^*(V',\sigma')$ be 
the homomorphism induced from the presymplectic map 
$V \to V'$. 
Then we see that $\varphi_S = \varphi_{S'}\circ \pi$, 
$\varphi_T = \varphi_{T'}\circ \pi$ and $\pi$ satisfies 
the weak approximation property in \cite{GMS} 
with respect to the states 
$\varphi_{S'}$, $\varphi_{T'}$ and therefore  
by \cite[Corollary~2.6]{GMS}
\[
(\varphi_S^{1/2}|\varphi_T^{1/2}) 
= (\varphi_{S'}^{1/2}| \varphi_{T'}^{1/2}) 
= \sqrt{\det\left( 
\frac{2\sqrt{A'B'}}{A'+B'}
\right)}.
\]

\begin{Theorem}\label{finite}
Let $(V,\sigma)$ be a finite-dimensional presymplectic 
vector space. For polarizations $S$ and $T$ of $(V,\sigma)$, 
we have 
\[
(\varphi_S^{1/2}|\varphi_T^{1/2}) 
= \sqrt{\det\left( 
\frac{2\sqrt{AB}}{A+B}
\right)},
\]
where positive quadratic forms $A$, $B$ are defined by 
\[
2A = (\sqrt{S} + \sqrt{\overline S})^2, 
\quad
2B = (\sqrt{T} + \sqrt{\overline T})^2.
\]
Notice that $2\sqrt{AB} \leq A + B$ 
(geometric mean is majorized by arithmetic mean) and 
the determinat in the above formula is always well-defined. 
\end{Theorem}

\subsection{Central Decomposition}
We shall decompose regular representations of 
$C^*(V,\sigma)$ 
with the help of Fourier transform on the central 
subalgebra $C^*(V_0,0)$, where $V_0 = \ker \sigma$. 

To this end, choose Lebesgue measures $dx$ on $V$ and  
$dx_0$ on $V_0$ respectively. A Lebesgue measure $d\dot x$ 
on the quotient space $V/V_0$ is then specified by the relation 
\[
\int_V f(x)\,dx = \int_{V/V_0} d\dot x 
\int_{V_0} f(x + x_0)\,dx_0. 
\] 

If $\pi$ is a regular representation of $C^*(V,\sigma)$ 
satisfying $\pi(e^{ix_0}) = e^{i\omega(x_0)} 1$ 
for $x_0 \in V_0$ with $\omega \in V_0^*$, then 
we have the integration identity for $f \in \sS(V)$
\[
\int_V f(x) \pi(e^{ix})\,dx =  
\int_{V/V_0} d\dot x 
\int_{V_0} dx_0\, f(x+x_0) \pi(e^{i(x+x_0)})
= \int_{V/V_0} d\dot x f_\omega(x) \pi(e^{ix}),
\]
where the function $f_\omega$ on $V$, which is defined by 
\[
f_\omega(x) 
= \int_{V_0} e^{i\omega(x_0)} f(x+x_0)\,dx_0,
\]
satisfies  
(i) $f_\omega(x+x_0) = f_\omega(x) e^{-i\omega(x_0)}$ 
for $x_0 \in V_0$ and  
(ii) $f_\omega$ is rapidly decreasing when restricted to 
a complementary subspace of $V_0$. 

Denote by $\sS(V/\omega)$ 
the totality of functions satisfying these two conditions. 
The integration 
$\int_{V/V_0} f(x) \pi(e^{ix})\, d\dot x$ is well-defined 
for $f \in \sS(V/\omega)$ and 
suggests the following *-algebra structure 
in $\sS(V/\omega)$: 
\[
(f*g)(z) = \int_{V/V_0} 
f(z') g(z-z') e^{-i\sigma(z',z)/2} 
d\dot z'
\]
from 
\[
\int_{V/V_0} f(x) \pi(e^{ix})d\dot x 
\int_{V/V_0} g(y) \pi(e^{iy}) d\dot y 
= \int_{V/V_0} (f*g)(z) \pi(e^{iz}) d\dot z
\]
and $f^*(x) = \overline{f(-x)}$ from 
\[
\left(
\int_{V/V_0} f(x) \pi(e^{ix}) d\dot x
\right)^* 
= 
\int_{V/V_0} f^*(x) \pi(e^{ix}) d\dot x. 
\] 

The *-algebra $\sS(V/\omega)$ is then 
a Hilbert algebra with respect to 
the inner product 
\[
(f|g)_\omega = \int_{V/V_0} \overline{f(x)} g(x)\, d\dot x
\] 
with the associated trace functional on $\sS(V/\omega)$ 
given by 
\[
\tau_\omega(f) = f(0) 
\quad
\text{for $f \in \sS(V/\omega)$.} 
\]
Note that $\tau_\omega(f^**g) = (f|g)_\omega$ for 
$f, g \in \sS(V/\omega)$. 
For $x, y, z \in V$ and $f \in \sS(V/\omega)$, 
the multiplier product formula 
\[
(e^{ix}fe^{iy})(z) 
= f(z-x-y) 
e^{i(\sigma(x,y) - \sigma(x,z) - \sigma(z,y))/2}, 
\quad 
\]
remains valid. 

It is now obvious to see that the partial Fourier transform 
\[
\sS(V) \ni f \mapsto 
\int_{V_0^*}^\oplus f_\omega\,d\omega 
\in \int_{V_0^*}^\oplus \sS(V/\omega)\,d\omega
\]
gives rise to a decomposition of 
the relevant Hilbert algebra:
\[
(f*g)_\omega = f_\omega * g_\omega, 
\quad
(f^*)_\omega = (f_\omega)^*, 
\quad
(f|g) = \int_{V_0^*} d\omega\, (f_\omega|g_\omega)_\omega
\]
for $f, g \in \sS(V)$. Here $d\omega$ is the measure 
on $V_0^*$ in the duality relation with $dx_0$: 
\[
\int_{V_0^*} d\omega \int_{V_0} dx_0 g(x_0) e^{i\omega(x_0)} 
= g(0), 
\quad 
\int_{V_0} dx_0 \int_{V_0^*} d\omega h(\omega) e^{i\omega(x_0)} 
= h(0)
\]
for $g \in \sS(V_0)$ and $h \in \sS(V_0^*)$. 

If we denote by 
$C^*_\text{reg}(V/\omega,\sigma)$ the C*-closure of 
$\sS(V/\omega)$, then the C*-algebra 
$C^*(V/\omega,\sigma)$ introduced in the previous section is 
a multiplier subalgebra of $C^*_\text{reg}(V/\omega,\sigma)$. 

The following is a consequence of standard Fourier analysis. 

\begin{Lemma}
We have a decomposition of 
the C*-algebra $C^*_\text{reg}(V,\sigma)$ 
into a continuous field 
$\{ C^*_\text{reg}(V/\omega,\sigma) \}_{\omega \in V_0^*}$ 
of C*-algebras so that any regular 
representation $\pi$ of $C^*(V,\sigma)$ is covariantly 
decomposed into the form 
\[
\pi = \int_{V_0^*}^\oplus \pi_\omega\,\nu(d\omega),
\]
where $\nu$ is a measure on $V_0^*$ and 
$\{ \pi_\omega \}$ is a $\nu$-measurable field of 
regular representations of $\{ C^*(V/\omega,\sigma) \}$.
\end{Lemma}

Given a polarization $S$ of $(V,\sigma)$, 
if one applies the above decomposition to 
the GNS-representation of the quasifree state $\varphi_S$ 
and compare it with Proposition~\ref{central}, 
an $L^1$-decomposition of $\varphi_S$ is obtained:
\[
\varphi_S = \int_{V_0^*} 
\varphi_{S,\omega}\,\nu_{S_0}(d\omega), 
\]
where $\nu_{S_0}$ is a gaussian measure of 
covariance form $S_0 = S|_{V_0^\C\times V_0^\C}$, 
$\varphi_{S,\omega}$ is the state of $C^*(V/\omega,\sigma)$ 
introduced in \S~2.3
and $\{ \varphi_{S,\omega} \}$ is 
a $\nu_{S_0}$-measurable family of states of 
$\{ C^*(V/\omega,\sigma) \}$. 

Now the results on density operators 
are rewritten in terms of the Hilbert algebra $\sS(V/\omega)$: 
The density function 
\[
\rho_{S,\omega}(x) = e^{-i\omega(x_0)} e^{-S(x_1,x_1)/2} 
\quad 
\text{for $x = x_0 + x_1 \in V_0 + V_1$},
\]
where $V_1 = \{ x \in V; S(x,V_0) = 0\}$, 
satisfies 
\[
\tau_\omega(\rho_{S,\omega}*f) 
= \int_{V/V_0} f(x) \varphi_{S,\omega}(e_\omega^{ix})\, 
d\dot x, 
\quad 
f \in \sS(V/\omega)
\]
and its square root is given by 
\[
\rho_{S,\omega}^{1/2}(x) 
= \frac{1}{\sqrt{N_{S_1}}} 
e^{-A(x_1,x_1)/2} e^{-i\omega(x_0)}
\quad
\text{with} 
\quad 
N_{S_1} = \int_{V_1} e^{-A(x_1,x_1)}\,dx_1.
\]

Given a polarization $S$ of a presymplectic vector space 
$(V,\sigma)$, let $\dot S$ be the polarization 
of the quotient symplectic vector space $(V/V_0,\dot\sigma)$  
defined by 
\[
\dot S(\dot x,\dot y) = S(x_1,y_1), 
\]
where $x_1$ (resp.~$y_1$) denotes 
the projection of $x \in V^\C$ (resp.~$y \in V^\C$) 
to the $(S+\overline S)$-orthogonal complement of 
$V_0^\C$. 

The follwing is immediate from the definition. 

\begin{Lemma}
For $x \in V^\C$, we have 
\begin{align*}
{\dot S}(\dot x) &= \inf \{ S(x+y_0); y_0 \in V_0^\C \},\\ 
({\dot S} + \overline{\dot S})(\dot x) 
&= \inf \{ (x+y_0,x+y_0)_S; y_0 \in V_0^\C \}
\end{align*}
and 
\[
{\dot A}(\dot x) = \inf \{ A(x+y_0); y_0 \in V_0^\C \}
\quad
\text{with}\ 
\dot A = ({\dot S} + \overline{\dot S})/2 
+ \sqrt{{\dot S}\overline{\dot S}}.
\] 
\end{Lemma}

Let $T$ be another polarization of $(V,\sigma)$ and 
let $F_0$ be the $(T+\overline T)$-orthogonal 
projection to the subspace $V_0^\C$.  

\begin{Proposition}
The transition amplitude 
$(\varphi_{S,\omega}^{1/2}|\varphi_{T,\omega}^{1/2})$ 
is given by 
\[
(\varphi_{S,\omega}^{1/2}|\varphi_{T,\omega}^{1/2})
= (\varphi_{\dot S}^{1/2}|\varphi_{\dot T}^{1/2})
e^{-(\dot A + \dot B)^{-1}(\Delta\omega)/2}. 
\]
Here $\Delta: V_0^* \to (V/V_0)^*$ is a linear map defined by 
\[
\langle \Delta\omega, \dot x \rangle 
= \omega((E_0-F_0)x), 
\ x \in V.
\]
\end{Proposition}

\begin{proof}
We use the expression 
\[
\overline{\rho_{S,\omega}^{1/2}}(x) 
\rho_{T,\omega}^{1/2}(x) = 
(N_{\dot S} N_{\dot T})^{-1/2} 
e^{-{\dot A}(x)/2 -{\dot B}(x)/2} 
e^{i\omega((E_0-F_0)x)} 
\]
in the following rewriting: 
\begin{align*}
(\varphi_{S,\omega}^{1/2}|\varphi_{T,\omega}^{1/2})
&= \tau_\omega(\rho_{S,\omega}^{1/2}*\rho_{S,\omega}^{1/2})
= \int_{V/V_0} \overline{\rho_{S,\omega}^{1/2}(x)} 
\rho_{T,\omega}^{1/2}(x) 
\, d\dot x\\
&= \frac{1}{\sqrt{N_{\dot S}N_{\dot T}}} 
\int_{V/V_0} e^{-(\dot A + \dot B)(\dot x)/2} 
e^{i\langle\Delta\omega,\dot x\rangle} \,d\dot x\\
&= \sqrt{
\det\left(
\frac{2\sqrt{\dot A\dot B}}{\dot A + \dot B}
\right)
}
e^{-(\dot A + \dot B)^{-1}(\Delta\omega)/2}\\ 
&= (\varphi_{\dot S}^{1/2}|\varphi_{\dot T}^{1/2})
e^{-(\dot A + \dot B)^{-1}(\Delta\omega)/2}.  
\end{align*}
\end{proof}

\section{Infinite-Dimensional Analysis}
From here on, we shall work with 
an infinie-dimensional 
presymplectic vector space $(V,\sigma)$. 

\subsection{Topological Equivalence on Polarizations}
\begin{Lemma}
Let $S,T \in \text{Pol}(V,\sigma)$ be such that 
we can find $0 \not= x \in V^\C$ and $\epsilon>0$ satisfying 
$(x,x)_T \not= 0$ and $(x,x)_S \leq \epsilon (x,x)_T$. Then 
$(\varphi_S^{1/2}|\varphi_T^{1/2}) \leq 2\epsilon^{1/4}$.  
\end{Lemma}

\begin{proof}
Consider the real subspace $W^\C = \C x + \C x^*$ 
and let $S_W$ and $T_W$ be the restrictions of $S$ and $T$ 
to $W^\C$. Set 
\[
2A_W = S_W + \overline{S_W} + 2\sqrt{S_W\overline{S_W}}, 
\quad
2B_W = T_W + \overline{T_W} + 2\sqrt{T_W\overline{T_W}}. 
\]
From the inequality 
$T_W + \overline{T_W} \geq 2 \sqrt{T_W \overline{T_W}}$, 
$\frac{1}{2} (x,x)_T \leq B_W(x,x) \leq (x,x)_T$ 
and similarly for $A_W$ and $(\ ,\ )_S$. 
We then have 
\[
\frac{1}{2} (x,x)_T \leq A_W(x,x) + B_W(x,x) 
\]
and 
\[
\sqrt{A_WB_W}(x,x) \leq 
A_W(x,x)^{1/2} B_W(x,x)^{1/2}
\leq \epsilon^{1/2} 
(x,x)_T^{1/2} (x,x)_T^{1/2},
\]
whence
\[
\frac{2\sqrt{A_WB_W}(x,x)}{(A_W+B_W)(x,x)} 
\leq 4\epsilon^{1/2}. 
\]
Let $C$ be the operator on $W^\C$ defined by 
\[
2\sqrt{A_WB_W}(\xi,\eta) = (A_W+B_W)(\xi,C\eta). 
\]
Then $0 \leq C \leq 1$ with respect to the inner product 
$A_W + B_W$ and the above inequality implies 
$\text{Sp}(C) \cap [0,4\epsilon^{1/2}] \not= \emptyset$ and 
therefore 
\[
\det\left(
\frac{2\sqrt{A_WB_W}}{A_W+B_W}
\right) 
\leq 4\epsilon^{1/2}.
\]
Now apply the increasing property of 
transition amplitude (\cite[Lemma~4.1]{GMS}) to get 
\[
(\varphi_S^{1/2}|\varphi_T^{1/2}) 
\leq (\varphi_{S_W}^{1/2}|\varphi_{T_W}^{1/2}) 
\leq 2\epsilon^{1/4}.
\]
\end{proof}

\begin{Corollary}
Let $S, T \in \text{Pol}(V,\sigma)$. 
Then $(\varphi_S^{1/2}|\varphi_T^{1/2}) = 0$ 
unless $(\ ,\ )_S$ and $(\ ,\ )_T$ are equivalent. 
\end{Corollary}

\subsection{Hilbert-Schmidt Estimates}

In what follows, assume that $(\ ,\ )_S$ and $(\ ,\ )_T$ 
are equivalent and complete on $V^\C$. 
Then we can find a bounded operator $R$ with a bounded 
inverse which is positive 
with respect to the inner product $(\ ,\ )_S$ 
($R$ being said to be $S$-positive) and satisfies 
\[
(x,y)_T = (Rx,Ry)_S, 
\quad x, y \in V^\C.
\]
In the notation of functional calculus (see Appendix~A), 
$
R = \left(
\frac{T + \overline{T}}{S + \overline{S}}
\right)^{1/2}$. 

Since the norm of a Laurent polynomial $f(R)$ of $R$ 
satisfies 
$\| f(R)\|_S = \| R^{-1} f(R)R\|_T = \| f(R)\|_T$,  
we shall omit the reference subscript. 

Let $\delta = 2\log(\| R\|\,\| R^{-1}\|)$, 
which is the projective distance between 
$(\ ,\ )_S$ and $(\ ,\ )_T$. 
Let $A_S$ and $B_S$ be $S$-positive operators 
representing $A$ and $B$ relative to 
the inner product $(\ ,\ )_S$. 
We also set 
$\bS = (S+\overline S)\backslash S$ and 
$\T = (T+\overline T)\backslash T$ as before. 

In the next lemma, 
Hilbert-Schmidt norm as well as operator norm is 
the one based on the inner product $(\ ,\ )_S$. 

\begin{Lemma}\label{AY}~ 
\begin{enumerate}
\item
\[
\| A_S - B_T \|_{HS} 
\leq 2\sqrt{2}(1 + e^{\delta/2}) 
\left\| 
\sqrt{\bS} - \sqrt{\T} 
\right\|_{HS}. 
\]
\item
\[
\| A_S - B_S \|_{HS} 
\leq 2\| 1 - R^2\|_{HS} 
+ 2\sqrt{2}(1 + e^{\delta/2}) 
\| R^2\|
\left\| 
\sqrt{\bS} - \sqrt{\T} 
\right\|_{HS}
\]
\item 
\[
\left\| \sqrt{\bS} - \sqrt{\T} \right\|_{HS} 
\leq \frac{e^{\pi/4}}{\sqrt{2}} 
e^\delta(1 + e^{\delta/2}) 
\| R^{-2}\|
\left\| A_S - B_S \right\|_{HS}
\] 
and 
\[
\| 1 - R^2\|_{HS} \leq 
e^{\delta/2}\| A_S - B_S\|_{HS} 
+ 2\sqrt{2} (e^{\delta} + e^{\delta/2}) 
\left\| \sqrt{\bS} - \sqrt{\T} \right\|_{HS}.
\]
\end{enumerate}
\end{Lemma}

\begin{proof}
Imitate the computations in \cite[Lemma~8.4]{AY} 
\end{proof}

\begin{Corollary}
The following two conditions are equivalent. 
\begin{enumerate}
\item 
The operator $\frac{A-B}{A+B}$ is in the Hilbert-Schmidt 
class. 
\item 
Both of $\frac{T+\overline{T}}{S+\overline{S}} - 1$ 
and $\left(\frac{S}{S+\overline{S}}\right)^{1/2} 
- \left(\frac{T}{T+\overline{T}}\right)^{1/2}$ 
are in the Hilbert-Schmidt class. 
\end{enumerate}
\end{Corollary}

\begin{proof}
Use the expression 
\[
A_S - B_S = \frac{A+B}{S + \overline S}
\frac{A + B}{A - B}
\]
and compute as follows 
(see Appendix~A for the first equality):  
\begin{align*}
\| A_S - B_S\|_{HS} 
&= \left\| 
\left( 
\frac{S + \overline S}{A + B}
\right)^{1/2} 
\frac{A - B}{S + \overline S} 
\left( 
\frac{S + \overline S}{A + B}
\right)^{-1/2}
\right\|_{(A+B)-HS}\\
&= 
\left\| 
\left( 
\frac{A+B}{S + \overline S}
\right)^{1/2} 
\frac{A - B}{A + B} 
\left( 
\frac{A + B}{S + \overline S}
\right)^{1/2}
\right\|_{(A+B)-HS}\\
&\leq 
\left\| 
\frac{A+B}{S + \overline S} 
\right\|\, 
\left\| 
\frac{A - B}{A + B} 
\right\|_{(A+B)-HS}\\ 
&\leq 
\| 1 + R^2\| 
\left\| 
\frac{A - B}{A + B} 
\right\|_{(A+B)-HS}.  
\end{align*}
Here at the last inequality, 
we have applied the operator inequality
$\frac{A+B}{S+\overline S} \leq 1+R^2$ 
relative to the inner product $S + \overline S$, 
which is a consequence of $A \leq S+\overline S$ and 
$B \leq T + \overline T$. 
\end{proof}

\subsection{Orthogonality of Quasifree States}

\begin{Lemma}
Given admissible polarizations $S$, $T$ 
on a complete presymplectic vector space $(V,\sigma)$, 
we can find a family of closed separable subspaces 
$\{ V_i \}_{i \in I}$ of $V$ such that 
$S(V_i,V_j) = 0 = T(V_i,V_j)$
if $i \not= j$ and 
$V = \overline{\oplus_{i \in I} V_i}$. 
\end{Lemma}

\begin{proof}
This follows from a standard maximality argument 
relying on Zorn's lemma (cf.~\cite[Lemma~6.9]{Ar}). 
\end{proof}

In view of this lemma, we may restrict ourselves to 
a separable $V$. 
Choose an increasing sequence of finite-dimensional 
vector space $\{ V_n \}_{n \geq 1}$ such that 
$\dim V_n = n$ and 
$V = \overline{\bigcup_{n \geq 1} V_n}$. 
Let $S_n = S|_{V_n^\C}$ and $T_n = T|_{V_n^\C}$ be 
the restrictions with 
$\bS_n = (S_n + \overline{S_n})\backslash S_n$ and 
$\T_n = (T_n + \overline{T_n})\backslash T_n$ be the associated 
linear operators on $V_n^\C$. 
Let $E_n$ (resp.~$F_n$) be the $(S+\overline S)$-orthogonal 
(resp.~$(T+\overline T)$-orthogonal) projection to 
the subspace $V_n^\C$. Then, for $x, y \in V^\C$, we have 
\[
E_n \bS E_n = \bS_n E_n, 
\quad 
F_n \T F_n = \T_n F_n
\] 
and hence 
\[
\bS^{1/2} = \lim_{n \to \infty} \bS_n^{1/2} E_n
\quad\text{and}\quad 
\T^{1/2} = \lim_{n \to \infty} \T_n^{1/2} F_n
\]
in the strong operator topology. 

Choose an $(S+\overline S)$-orthonormal basis 
$\{ \xi_n\}_{n \geq 1}$ of 
$V^\C$ so that $V_n^\C = \C \xi_1 + \dots + \C \xi_n$. 
Then, for $1 \leq m \leq n$, 
\[
\sum_{j=1}^m \| (\bS_n^{1/2}E_n - \T_n^{1/2}F_n)\xi_j\|_S^2 
= \sum_{j=1}^m \| (\bS_n^{1/2} - \T_n^{1/2})\xi_j \|_S^2 
\leq \sum_{j=1}^n \| (\bS_n^{1/2} - \T_n^{1/2})\xi_j\|_S^2. 
\]
Taking $n \to \infty$ and then $m \to \infty$, we have 
\[
\| \bS^{1/2} - \T^{1/2}\|_{HS}^2 \leq 
\liminf_{n \to \infty} 
\| \bS_n^{1/2} - \T_n^{1/2}\|_{HS}^2. 
\]

Next, for $x, y \in V^\C$, 
$E_nR^2E_n = E_nR_n^2E_n = R_n^2E_n$ 
with $R_n = (T_n + \overline{T_n})/(S_n + \overline{S_n})$ 
implies 
\[
R = \lim_{n \to \infty} (E_nR^2E_n)^{1/2} 
= \lim_{n \to \infty} (R_n^2E_n)^{1/2} 
= \lim_{n \to \infty} R_n E_n
\]
in the strong operator topology. As a result, we have 
\[
\| R - 1\|_{HS} \leq \liminf_{n \to \infty} 
\| R_n - 1_n \|_{HS}.
\]

\begin{Lemma}
Unless both of $\bS^{1/2} - \T^{1/2}$ and 
$R-1$ are in the Hilbert-Schmidt class, we have 
$(\varphi_S^{1/2}|\varphi_T^{1/2}) = 0$.   
\end{Lemma}

\begin{proof}
Let $2A_n = (S_n^{1/2} + {\overline S}_n^{1/2})^2$, 
$2B_n = (T_n^{1/2} + {\overline T}_n^{1/2})^2$ and set
$C_n = \frac{A_n-B_n}{A_n+B_n}$. Then 
$-1 \leq C_n \leq 1$ and 
\[
\det\left(
\frac{2\sqrt{A_nB_n}}{A_n+B_n}\right) 
= \sqrt{\det(1-C_n^2)}. 
\]

From the estimate discussed above, 
the assumption implies 
\[
\liminf_{n \to \infty} 
\| \bS_n^{1/2} - \T_n^{1/2} \|_{HS} 
= +\infty 
\quad 
\text{or}
\quad 
\liminf_{n \to \infty} \| R_n - 1_n \|_{HS} = +\infty, 
\]
which, in turn, gives 
\[
\liminf_{n \to \infty} 
\left\| 
\frac{A_n}{S_n + \overline{S_n}} 
- \frac{B_n}{S_n + \overline{S_n}} 
\right\|_{HS} = +\infty
\]
by the inequalities in Lemma~\ref{AY}. 
Since 
$\| 1_n + R_n^2\| = 1 + \| R_n^2\| \leq 1 + \| R^2\|$, 
the inequality 
\begin{align*}
\left\| 
\frac{A_n - B_n}{S_n + \overline{S_n}} 
\right\|_{HS} 
&\leq 
\| 1_n + R_n^2\|\,
\left\| 
\frac{A_n-B_n}{A_n+B_n} 
\right\|_{(A_n+B_n)-HS}\\
&\leq (1 + \| R\|^2) 
\left\| 
\frac{A_n-B_n}{A_n+B_n} 
\right\|_{(A_n+B_n)-HS}
\end{align*}
implies 
\[
\liminf_{n \to \infty} 
\text{tr} 
\left( 
\frac{A_n - B_n}{A_n + B_n}
\right)^2 = +\infty, 
\] 
whence the equality 
$(\varphi_S^{1/2}|\varphi_T^{1/2}) = 
\lim_{n \to \infty} (\varphi_{S_n}^{1/2}|\varphi_{T_n}^{1/2})$ 
(\cite[Theorem~4.3]{GMS}) is used to have  
\begin{align*}
(\varphi_S^{1/2}|\varphi_T^{1/2}) 
&= \lim_{n \to \infty} 
(\varphi_{S_n}^{1/2}|\varphi_{T_n}^{1/2}) 
= \lim_{n \to \infty} 
\det\left(
\frac{2\sqrt{A_nB_n}}{A_n+B_n} 
\right)\\
&= \lim_{n \to \infty} 
\sqrt{\det(1-C_n)}
= \lim_{n \to \infty} 
\exp\left(\frac{1}{2} 
\text{tr} \log(1-C_n^2) 
\right)\\ 
&\leq \limsup_{n \to \infty} 
\exp(-\frac{1}{2}\text{tr}(C_n^2))
= \exp\left(
-\frac{1}{2} \liminf_{n \to \infty} \text{tr}(C_n^2)
\right)
 = 0.
\end{align*}
\end{proof}

\section{Quadrature on Polarizations}
\subsection{Quadrature on Polarizations}
Start with a presymplectic vector space 
$(V,\sigma)$ such that $V^\C$ is a Hilbert space 
relative to an inner product of the form 
$S + \overline{S}$ with $S$ a polarization of $(V,\sigma)$
(recall that 
a polarization possessing this property is said to be 
admissible). 
Note that any Hilbert space inner product on $V^\C$ is 
unique up to metrical equivalence and represents 
$\sigma$ by a bounded operator. 
We shall now review the construction 
of phase-space doubling 
(see \cite{AS,V} for example).

Consider a presymplectic vector space of the form 
$(V \oplus V, \sigma \oplus -\sigma)$. 
Given an admissible polarization $S$, define 
a positive form $P$ on $V^\C \oplus V^\C$ by 
\begin{align*}
P(x \oplus y,x'\oplus y') 
&= 
\begin{pmatrix}
x^* & y^*
\end{pmatrix}
\begin{pmatrix}
S & \sqrt{S{\overline S}}\\
\sqrt{S{\overline S}} & {\overline S}
\end{pmatrix}
\begin{pmatrix}
x'\\
y'
\end{pmatrix}\\
&= S(x,x') + \sqrt{S\overline{S}}(x,y') 
+ \sqrt{S\overline{S}}(y,x') 
+ \overline{S}(y,y'), 
\end{align*}
which is a polarization of 
$(V\oplus V, \sigma \oplus -\sigma)$. 

To look into the topology induced from $P + \overline P$, 
we consider the `rotation' of 
$(V\oplus V, \sigma \oplus -\sigma)$ by an angle $\pi/4$. 
Let $R_{\pi/4}: V\oplus V \to V \oplus V$ be defined by 
$R_{\pi/4}(x\oplus y) = 
\frac{x-y}{\sqrt{2}} \oplus 
\frac{x+y}{\sqrt{2}}$. 
Then the identities 
\[
P_{\pi/4} \equiv 
R_{\pi/4}^{-1} P R_{\pi/4} 
= \frac{1}{2} 
\begin{pmatrix}
(S^{1/2} + {\overline S}^{1/2})^2 & {\overline S} - S\\
{\overline S} - S & (S^{1/2} - {\overline S}^{1/2})^2
\end{pmatrix}
\]
and 
\[
(\sigma \oplus -\sigma)_{\pi/4}
\equiv 
R_{\pi/4}^{-1} 
\begin{pmatrix}
\sigma & 0\\
0 & -\sigma
\end{pmatrix}
R_{\pi/4} 
= 
\begin{pmatrix}
0 & -\sigma\\
-\sigma & 0
\end{pmatrix}
\]
show that $R_{\pi/4}$ is a presymplectic isomorphism 
of $(V\oplus V,(\sigma \oplus -\sigma)_{\pi/4})$ onto 
$(V\oplus V, \sigma \oplus -\sigma)$ and 
$P(R_{\pi/4}\xi,R_{\pi/4}\eta) = P_{\pi/4}(\xi,\eta)$ 
for $\xi, \eta \in V^\C \oplus V^\C$. 

The kernel of $P + \overline P$ corresponds to 
that of 
\[
P_{\pi/4} + \overline{P_{\pi/4}} 
= 
\begin{pmatrix}
(S^{1/2} + {\overline S}^{1/2})^2 & 0\\
0 & (S^{1/2} - {\overline S}^{1/2})^2
\end{pmatrix} 
\]
by the presymplectic isomorphism $R_{\pi/4}$, 
which is equal to $0 \oplus (\ker\sigma)^\C$ 
(cf.~$(\bS^{1/2} - {\overline\bS}^{1/2})^2 
= (\bS^{1/2} +{\overline\bS}^{1/2})^{-2} 
(\bS - {\overline\bS})^2$). 
Thus 
\[
\ker (\ ,\ )_P = 
\{ x \oplus -x; x \in (\ker\sigma)^\C \}
\]
and $P$ induces a non-degenerate polarization on the 
quotient presymplectic vector space 
$(V \oplus V)/ \{ x \oplus -x; x \in \ker\sigma \}$. 

Clearly the quotient by the kernel of 
$P_{\pi/4} + \overline{P_{\pi/4}}$ is equal to 
$V \oplus {\dot V}$ with ${\dot V} = V/\ker\sigma$ and 
$R_{\pi/4}$ induces a presymplectic isomorphism 
$V \oplus {\dot V} \to (V\oplus V)/\{(x\oplus -x); 
x \in \ker\sigma \}$. 

Since $S + \overline S \leq 
(S^{1/2} + {\overline S}^{1/2})^2 \leq 2(S + \overline S)$, 
the topology induced from $P_{\pi/4} + \overline{P_{\pi/4}}$ 
is hilbertian when restricted to $V^\C \oplus 0$, 
while the topology on $0 \oplus V^\C$ is associated to 
the positive form $(S^{1/2} - {\overline S}^{1/2})^2$,  
which is generaly different from 
that of $V^\C$ even for a non-degenerate $\sigma$. 

\begin{Lemma}
Given an admissible polarization $T$ of a hilbertian 
presymplectic vector space $(V,\sigma)$, let 
$W_T$ be the Hilbert space associated to 
the positive form $(T^{1/2} - {\overline T}^{1/2})^2$ 
($W_T$ being therefore a completion of the quotient space 
${\dot V} = V/\ker\sigma$). 
Then we have an isometry $U_T: W_T^\C \to V^\C$ 
($W_T^\C$ and $V^\C$ being furnished with inner products 
$(T^{1/2} - {\overline T}^{1/2})^2$ and 
$(T^{1/2} + {\overline T}^{1/2})^2$ respectively) 
such that 
\[
U_T{\dot x} = 
\frac{T^{1/2} - {\overline T}^{1/2}}
{T^{1/2} + {\overline T}^{1/2}}x 
\]
for ${\dot x} \in {\dot V}^\C$ with $x \in V^\C$.  

Furthermore, the composition 
${\dot U}_T: W_T^\C \to V^\C \to {\dot V}^\C$ is unitary 
if ${\dot V}^\C$ is furnished with 
the inner product 
$({\dot T}^{1/2} + {\overline{\dot T}}^{1/2})^2$. 
Here the positive form $\dot T$ on $\dot V$ is defined by 
${\dot T}(\dot x, \dot y) = T(x,y)$ with 
representatives $x$ and $y$ taken from 
the $(T+\overline T)$-orthogonal complement of $V_0^\C$. 
Note that 
${\dot T}(\dot x,\dot x) 
= \inf\{ T(x,x); \dot x = x + V_0^\C\}$. 
\end{Lemma}

\begin{proof}
By passing to the quotient $\dot V$, we may assume 
that $\ker\sigma = \{ 0\}$. 

Let $\T = (T + \overline T)\backslash T$ 
be a $(T + \overline T)$-positive ratio operator, 
which satisfies the inequality
$1 \leq \sqrt{\T} + \sqrt{1-\T} \leq \sqrt{2}$ 
and hence $\sqrt{\T} + \sqrt{1-\T}$ has a bounded inverse. 
The identity 
\begin{align*}
(T^{1/2} + {\overline T}^{1/2})^2
&\left(
\frac{\sqrt{\T} - \sqrt{1-\T}}
{\sqrt{\T} + \sqrt{1-\T}}
x, 
\frac{\sqrt{\T} - \sqrt{1-\T}}
{\sqrt{\T} + \sqrt{1-\T}}
y
\right)\\
&= \left(
\frac{\sqrt{\T} - \sqrt{1-\T}}
{\sqrt{\T} + \sqrt{1-\T}}
x, 
(\sqrt{\T} + \sqrt{1-\T})^2
\frac{\sqrt{\T} - \sqrt{1-\T}}
{\sqrt{\T} + \sqrt{1-\T}}
y
\right)_T\\
&= (x, (\sqrt{\T} - \sqrt{1-\T})^2y)_T
= (T^{1/2} - {\overline T}^{1/2})^2(x,y)
\end{align*}
then shows that we have an isometry $U: W^\C \to V^\C$ such 
that 
\[
Ux = 
\frac{T^{1/2} - {\overline T}^{1/2}}
{T^{1/2} + {\overline T}^{1/2}}x
\quad 
\text{for $x \in V^\C \subset W^\C$.}
\]
\end{proof}

\begin{Lemma}
Let $S$ and $T$ be admissible polarizations of 
a hilbertian presymplectic vector space $(V,\sigma)$. 
Then positive forms 
$(S^{1/2} - {\overline S}^{1/2})^2$,  
$(T^{1/2} - {\overline T}^{1/2})^2$ on $V^\C$ are equivalent 
and therefore $W_S = W_T$. The common hilbertian space is 
denoted by $W$.  

Moreover, we have 
\[
\frac{(S^{1/2} - {\overline S}^{1/2})^2}
{(T^{1/2} - {\overline T}^{1/2})^2} 
= U_T^* 
\frac{(T^{1/2} + {\overline T}^{1/2})^2}
{(S^{1/2} + {\overline S}^{1/2})^2}
U_T. 
\]
Note that the left hand side is a ratio operator defined 
on the hilbertian space $W^\C$. 
\end{Lemma} 

\begin{proof}
Keep the notation in the proof of the previous lemma. 
Since $\ker\sigma = \{ 0\} = 
\ker(\sqrt{\T} - \sqrt{1-\T})$ by assumption, 
$U$ has a dense range and hence it is a unitary map. 
For $x \in V^\C$ and $y = Uy'$ with $y' \in V^\C$, 
we see
\begin{align*}
(T^{1/2} - {\overline T}^{1/2})^2(x,y) 
&= (T^{1/2} - {\overline T}^{1/2})^2( 
\frac{T^{1/2} - {\overline T}^{1/2}}
{T^{1/2} + {\overline T}^{1/2}}x,y')\\
&= (T^{1/2} - {\overline T}^{1/2})^2( 
\frac{T^{1/2} - {\overline T}^{1/2}}
{T^{1/2} + {\overline T}^{1/2}}x,U^*y). 
\end{align*}
Since both of the initial and final expressions are
continuous in $y \in W^\C$, we have 
\[
(T^{1/2} - {\overline T}^{1/2})^2(x,y) 
= (T^{1/2} - {\overline T}^{1/2})^2( 
\frac{T^{1/2} - {\overline T}^{1/2}}
{T^{1/2} + {\overline T}^{1/2}}x,U^*y) 
\]
for $x \in V^\C$ and $y \in W^\C$. 

Taking the relation $S - \overline S = T - \overline T$ 
into account, we have the following for 
$y \in V^\C \subset W^\C$ and $x = Ux'$ with $x' \in V^\C$: 
\begin{align*}
(S^{1/2} - {\overline S}^{1/2})^2(x,y)
&= 
(S - \overline S)(x, 
\frac{S - \overline S}{(S^{1/2} + {\overline S}^{1/2})^2}y)\\
&= 
(T - \overline T)(x, 
\frac{T - \overline T}{(S^{1/2} + {\overline S}^{1/2})^2}y)\\
&= 
(T - \overline T)(x, 
\frac{(T^{1/2} + {\overline T}^{1/2})^2}
{(S^{1/2} + {\overline S}^{1/2})^2}
\frac{T - \overline T}{(T^{1/2} + {\overline T}^{1/2})^2}y)\\
&= 
(T^{1/2} - {\overline T}^{1/2})^2(x', 
\frac{(T^{1/2} + {\overline T}^{1/2})^2}
{(S^{1/2} + {\overline S}^{1/2})^2}
Uy)\\
&= 
(T^{1/2} - {\overline T}^{1/2})^2(
\frac{T^{1/2} - {\overline T}^{1/2}}
{T^{1/2} + {\overline T}^{1/2}}
x', U^*
\frac{(T^{1/2} + {\overline T}^{1/2})^2}
{(S^{1/2} + {\overline S}^{1/2})^2}
Uy)\\
&= 
(T^{1/2} - {\overline T}^{1/2})^2(x, U^*
\frac{(T^{1/2} + {\overline T}^{1/2})^2}
{(S^{1/2} + {\overline S}^{1/2})^2}
Uy). 
\end{align*}
Since $\frac{T^{1/2} - {\overline T}^{1/2}}
{T^{1/2} + {\overline T}^{1/2}}V^\C$ as well as 
$V^\C$ is dense in $W^\C$, we obtain the equivalence 
of positive forms
$(S^{1/2} - {\overline S}^{1/2})^2$,
$(T^{1/2} - {\overline T}^{1/2})^2$ on $W^\C$ and 
the equality  
\[
\frac{(S^{1/2} - {\overline S}^{1/2})^2}
{(T^{1/2} - {\overline T}^{1/2})^2} 
= U^* 
\frac{(T^{1/2} + {\overline T}^{1/2})^2}
{(S^{1/2} + {\overline S}^{1/2})^2}
U
\]
holds at the same time. 
\end{proof}

\begin{Corollary}[{\cite[Lemma~6.1]{AS}}]
The topology on 
$(V \oplus V)/\{ x\oplus -x; x \in \ker\sigma \}$ 
induced from $P + \overline P$ does not depend on 
the choice of an admissible polarization $S$. 
Let $\widehat V$ be its hilbertian completion and 
${\widehat V}_{\pi/4}$ be  
the rotated space of $\widehat V$. 
Then we have 
${\widehat V}_{\pi/4} 
= V \oplus W$. 
\end{Corollary}

Let $\widehat \sigma$ be the presymplectic form 
on $\widehat V$ induced from $\sigma \oplus -\sigma$. 
We regard $P$ as defining 
an admissible polarization of the presymplectic vector space 
$(\widehat V, \widehat\sigma)$ and call it 
the \textbf{quadrature} of $S$. 

\begin{Lemma}[{\cite[Lemma~5.8]{AS}}]
The spectrum of 
the ratio operator $(P+\overline P)\backslash P$ is a subset of 
$\{ 0, 1/2, 1\}$ with its spectral subspaces given by 
closures of 
\[
\{ [-\sqrt{1-\bS}x \oplus \sqrt{\bS}x]; x \in V^\C \}, 
\{ [x \oplus x]; x \in V_0^\C \}, 
\{ [\sqrt{\bS}x \oplus - \sqrt{1-\bS}x]; x \in V^\C \}
\]
respectively,  
where $[x\oplus y]$ denotes the quotient of 
$x \oplus y \in V^\C \oplus V^\C$ with respect to 
the subspace $\{ z \oplus - z; z \in (\ker\sigma)^\C \}$. 
\end{Lemma}

\begin{Lemma}[{\cite[\S 3]{Ar}}]\label{fock}
Let $y \in V \cap \ker(\bS(1-\bS)(2\bS - 1))^\perp$ 
be entirely analytic 
for $\{ e^{itH} \}_{t \in \R}$ (cf.~Proposition~\ref{KMS}). 
Then 
\[
e^{i[0\oplus y]}\,\varphi_P^{1/2} 
= e^{i[e^{H/2}y\oplus 0]}\, \varphi_P^{1/2}.
\]
Here $(e^{H/2}y)^* = e^{-H/2}y$ and the right hand side 
is, by definition, 
\[
\sum_{n=0}^\infty \frac{1}{n!} 
[e^{H/2}y\oplus 0]^n\,\varphi_P^{1/2},
\]
which is norm-convergent 
in the Hilbert space 
$\overline{C^*(\widehat V,\widehat\sigma)\,
\varphi_P^{1/2}}$. 
\end{Lemma}

\begin{Proposition}
Let $S$ be an admissible polarization.
Then we have the unitary map 
\[
\overline{C^*(V,\sigma)\varphi_S^{1/2} C^*(V,\sigma)} 
\to \overline{C^*(\widehat V,\widehat\sigma)\varphi_P^{1/2}}
\] 
defined by 
\[
e^{ix}\varphi_S^{1/2} e^{iy} \mapsto 
e^{i[x\oplus y]} \varphi_P^{1/2}, 
\quad x, y \in V.
\]
\end{Proposition}

\begin{proof}
Since the decomposition
\[
V^\C = (\ker\bS + \ker(1-\bS)) \oplus 
\ker(2\bS - 1) \oplus V_r^\C
\]
with $V_r^\C$ the orthogonal complement of 
$\ker(\bS(1-\bS)(2\bS - 1))$ gives rise to 
tensor product factorizations of 
$C^*(V,\sigma)$ and $\varphi_S$, we are reduced to 
checking each case separately. 

For $V_F^\C = \ker \bS + \ker(1-\bS)$, 
the associated representation is a Fock representation and 
the assertion follows from 
\[
C^*(V_F,\sigma_F) \varphi_{S_F}^{1/2} 
C^*(V_F,\sigma_F) 
\cong 
C^*(V_F,\sigma_F) \varphi_{S_F}^{1/2}\otimes 
\varphi_{S_F}^{1/2} C^*(V_F,\sigma_F). 
\]

Since the part $\ker(2\bS-1)$ produces 
a commutative algebra, we have 
\[
e^{ix}\varphi_S^{1/2} e^{iy} 
= e^{i(x+y)} \varphi_S^{1/2},
\]
whereas $[x\oplus y] = [x+y\oplus 0]$ for 
$x, y \in \ker(2\bS-1)$ shows that 
the correspondence in question is isometric. 

Finally assume that $\ker(\bS(1-\bS)(2\bS-1)) = \{ 0\}$. 
Then, by the previous lemma and the modular relation, 
the isometricity follows from 
\begin{align*}
(\varphi_P^{1/2}|e^{i[x\oplus y]}\varphi_P^{1/2}) 
&= (\varphi_P^{1/2}|e^{i[x\oplus 0]} 
e^{i[0\oplus y]}\varphi_P^{1/2})
= (\varphi_P^{1/2}|e^{i[x\oplus 0]} 
e^{i[e^{H/2}y\oplus 0]}\varphi_P^{1/2})\\ 
&= (\varphi_S^{1/2}| e^{ix} e^{ie^{H/2}y} 
\varphi_S^{1/2})
= (\varphi_S^{1/2}| e^{ix} \varphi_S^{1/2} e^{iy}).
\end{align*}
\end{proof}

\begin{Remark}
An antiunitary involution $J$ (the modular conjugation 
of $\varphi_S$) 
is defined on the Hilbert space 
$\overline{C^*(\widehat V,\widehat\sigma) \varphi_P^{1/2}}$
by 
\[
J(e^{i[x\oplus y]} \varphi_P^{1/2}) 
= e^{-i[y\oplus x]} \varphi_P^{1/2}, 
\quad 
x, y \in V.
\]
\end{Remark}

\subsection{Quadrate Polarizations}

\begin{Definition}
An admissible polarization $P$ of a complete 
presymplectic vector space 
is said to be \textbf{quadrate} if the spectrum of 
$\frac{P}{P+\overline P}$ is included in 
$\{ 0, \frac{1}{2}, 1 \}$. 
\end{Definition}

Let $(V,\sigma)$ be a hilbertian presymplectic vector space 
and set $V_0 = \ker\sigma$. 
Let $P$ and $Q$ be admissible quadrate polarizations 
of $(V,\sigma)$
with $P_0$ and $Q_0$ the restriction of 
$P$ and $Q$ to the subsapce $V_0^\C$ respectively. 

Let $E_0$ be the orthogonal projection to the subspace 
$V_0^\C$ with respect to the inner product $P+\overline P$. 
According to the decomposition $V^\C = V_0^\C + (1-E_0)V^\C$, 
operators $(P+\overline P)\backslash P$ and 
$(Q+\overline Q)\backslash Q$ are represented 
in the following block form 
\[
\frac{P}{P+\overline P} = 
\begin{pmatrix}
1/2 & 0 \\
0 & E
\end{pmatrix}, 
\qquad 
\frac{Q}{Q+\overline Q} = 
\begin{pmatrix}
1/2 & f\\
0 & F
\end{pmatrix}. 
\]

From the identity 
$(Q+\overline Q)\backslash Q + 
(Q+\overline Q)\backslash \overline Q = 1$, we have 
\[
F + \overline F = 1 - E_0 = E + \overline E, 
\quad 
f + \overline f = 0.
\]

Let $F_0$ and $F_1$ be the spectral projections of 
$(Q+\overline Q)\backslash Q$ associated to eigenvalues 
$1/2$ and $1$ respectively. Then 
\[
F_0 = 
\begin{pmatrix}
1 & 2f(\overline F - F)\\
0 & 0
\end{pmatrix}, 
\qquad 
F_1 = 
\begin{pmatrix}
0 & 2fF\\
0 & F
\end{pmatrix}
\]
and consequently 
\[
F^2 = F\ (\iff F_1^2 = F_1), 
\qquad 
F\overline F = 0 = \overline F F\ 
(\iff F_1\overline{F_1} = 0 = \overline{F_1} F_1).
\]
Note here that $F$ and $\overline F$ are 
\textit{not} necessarily 
orthogonal relative to the inner product $P+\overline P$. 

\begin{Lemma}[{\cite[Lemma~5.2]{Ar}}] We have the following. 
  \begin{enumerate}
  \item 
$(E-F)^2 = E\overline F E + \overline E F \overline E 
= F \overline E F + \overline F E \overline F$ 
is negative with respect to the inner product 
$P + \overline P$. 
\item 
$\overline{E-F} = -(E-F)$. 
\item
$[E,(E-F)^2] = 0 = [F,(E-F)^2]$. 
  \end{enumerate}
\end{Lemma}

By a computation similar to the proof of this lemma, 
we also have 

\begin{Lemma}
When $f = 0$, $(E-F)^2$ is negative with repsect to 
$Q + \overline Q$ as well and, if we denote the kernel 
projection of $(E-F)^2$ within $(1-E_0)V^\C$ 
by $c$ ($cE_0 = E_0c = 0$), then $\overline{c} = c$ and 
\[
P(cx,(1-c)y) = 0 = Q(cx,(1-c)y), 
\quad 
P(cx,cy) = Q(cx,cy)
\]
for $x, y \in V^\C$. 
\end{Lemma}

\subsection{Hilbert-Schmidt Approximations}

From here on, $(P+\overline P)\backslash P - 
(Q+\overline Q)\backslash Q$ and 
$(P+\overline P)\backslash (Q+\overline Q) - 1$ 
are assumed to be in the Hilbert-Schmidt class. 

\begin{Lemma}
We can find 
an increasing sequence of finite-dimensional subspaces 
$\{ V_n \}_{n \geq 1}$ 
with $\cup_{n \geq 1} V_n$ dense in $V$ 
and 
a sequence of admissible quadrate polarizations 
$\{ Q_n \}_{n \geq 1}$ such that 
\begin{enumerate}
\item 
if we denote by $W_n = V \ominus V_n$ 
the $(P+\overline P)$-orthogonal complement of $V_n$, 
then 
$Q_n(V_n^\C,W_n^\C) = 0$, 
\[
Q_n|_{V_n^\C\times V_n^\C} = Q|_{ V_n^\C\times V_n^\C}
\quad\text{and}\quad
Q_n|_{W_n^\C\times W_n^\C} = P|_{ W_n^\C\times W_n^\C}.
\] 
\item 
$\| Q_n - Q\|_{HS} \to 0$ as $n \to \infty$. 
\end{enumerate}
\end{Lemma}

\begin{proof}
Given a bounded operator $\gamma$ on $V^\C$ such that 
$\overline{\gamma} = \gamma$ and $\gamma = E_0\gamma(1-E_0)$, 
\[
\Gamma = e^\gamma = 
\begin{pmatrix}
E_0 & \gamma\\
0 &1-E_0
\end{pmatrix}
\]
defines a presymplectic transformation of $(V,\sigma)$ 
and the composition $Q\Gamma$ 
gives an admissible polarization.  

If we set $\gamma = -2f(1-2F) = E_0 - F_0$ 
in the expression 
\[
\frac{Q\Gamma}{Q\Gamma + \overline{Q\Gamma}} 
= 
\begin{pmatrix}
1/2 & f + \gamma(1-2F)/2\\
0 & F
\end{pmatrix}, 
\]
then we see that $Q' = Q\Gamma$ satisfies 
\[
\frac{Q'}{Q' + \overline{Q'}} 
= 
\begin{pmatrix}
1/2 & 0\\
0 & F
\end{pmatrix}.
\]

Choose an increasing sequence $\{ h_n\}_{n \geq 1}$ 
of projections so that
(i) $h_n$ is orthogonal relative to $P+\overline P$ and 
$Q' + \overline{Q'}$, 
(ii) $h_n$ commutes with $E$ and $F$,  
(iii) $h_n$ is of finite rank, 
(iv) $\overline{h_n} = h_n$ and 
(v) $\lim_{n \to \infty} h_n = 1-E_0$. 
(Recall that $(E-F)^2$ is in the Hilbert-Schmidt class 
and $P_c = Q_c$ on the kernel of $(E-F)^2$, 
see Lemma~5.9)

Then $fh_nV^\C$ is an increasing sequence of 
finite-dimensional *-invariant subspaces of $V_0^\C$ and 
we can find an increasing sequence 
$\{ g_n\}_{n \geq 1}$ of $(P+\overline P)$-orthogonal 
projections such that 
(i) $g_n$ is of finite rank, (ii) $\overline{g_n} = g_n$, 
(iii) $\lim_{n \to \infty} g_n = E_0$ and 
(iv) $g_n fh_n = fh_n$. 

Let $e_n = g_n+h_n$. Then $e_n$ is 
a $(P+\overline P)$-orthogonal projection such that 
(i) $e_n$ is of finite rank, (ii) $\overline{e_n} = e_n$ and 
(iii) $\lim_{n \to \infty} e_n = 1$. 

Define an admissible polarization of 
$({\widehat V}, {\widehat \sigma})$ by 
\[
Q_n'(x,y) = Q'(e_nx,e_ny) + P((1-e_n)x,(1-e_n)y). 
\]
In view of the fact that $h_n$ is 
$(Q'+\overline{Q'})$-orthogonal, 
we see $Q_n'(x,y) = 0$ 
for $x \in E_0V^\C$ and $y \in (1-E_0)V^\C$. 

Noticing that 
$e^{g_n(F_0-E_0)h_n}$ is a presymplectic transformation 
due to $\overline{g_n(F_0-E_0)h_n} = g_n(F_0-E_0)h_n$, 
we finally introduce an admissible polarization by 
$Q_n = Q_n' e^{g_n(F_0-E_0)h_n}$. 
From $(1-E_0)e_n = h_n$ and $[F,e_n] = 0$, we see that 
$(F_0-E_0)e_n = e_n(F_0-E_0)e_n$ and hence 
\[
e^{F_0-E_0}e_n = e^{g_n(F_0-E_0)h_n}e_n 
= e_n e^{g_n(F_0-E_0)h_n} 
= e_ne^{F_0-E_0}e_n,
\]
which is used to identify restrictions of $Q_n$: 
the results are 
\begin{align*}
Q_n(e_nx,e_ny) &= Q(e_nx,e_ny),\\ 
Q_n(e_nx, (1-e_n)y) &= 0,\\ 
Q_n((1-e_n)x,(1-e_n)y) 
&= P((1-e_n)x,(1-e_n)y).
\end{align*}

In this way, we have checked 
\[
\frac{Q_n}{P+\overline P} 
= e_n\frac{Q}{P+\overline P}e_n 
+ (1-e_n)\frac{P}{P+\overline P}(1-e_n) 
\]
and therefore 
\begin{multline*}
\frac{Q_n}{P+\overline P} - \frac{Q}{P+\overline P} 
=\\ 
(1-e_n)\frac{Q-P}{P+\overline P}e_n 
+ e_n\frac{Q-P}{P+\overline P}(1-e_n) 
+ (1-e_n)\frac{P-Q}{P+\overline P}(1-e_n)
\end{multline*}
converges to $0$ in the Hilbert-Schmidt norm 
because 
\[
\frac{Q-P}{P+\overline P} 
= \left( 
\frac{Q+\overline Q}{P+\overline P} - 1
\right)
\frac{Q}{Q+\overline Q} + 
\left(
\frac{Q}{Q + \overline Q} - \frac{P}{P+\overline P}
\right)
\]
is in the Hilbert-Schmidt class. 
\end{proof}

Choose an auxiliary Hilbert-Schmidt operator 
$\Theta: L \to V_0$ so that 
$\Theta L$ contains 
\[
(E_0-F_0)V^\C \cup \bigcup_{n \geq 1} (V_n \cap V_0). 
\] 
This is possible because $E_0-F_0$ is in the Hilbert-Schmidt 
class and $\dim V_n < +\infty$. 
Take $\omega \in (\Theta L)^*$ and consider 
quasifree states on $C^*(V/\omega,\sigma)$ as in \S 2.3. 

\begin{Lemma}
We have 
\[
\frac{Q_n}{Q_n + \overline{Q_n}} 
= 
\begin{pmatrix}
1/2 & fh_n\\
0 & F_n
\end{pmatrix}
\]
with $F_n = h_nF + (1-h_n)E$
and the equality  
$\varphi_{Q,\omega} = \varphi_{Q_n,\omega}$ holds 
on the C*-subalgebra $C^*(V_n/\omega_n,\sigma_n)$  
($\omega_n = \omega|_{V_n \cap V_0}$). 
\end{Lemma}

\begin{proof}
Since $e_n$ commutes with 
$(Q'+\overline{Q'})\backslash Q'$ and 
$(P+\overline{P})\backslash P$, we have 
\begin{align*}
Q_n'(x,y) &= 
(e_nx,\frac{Q'}{Q' + \overline{Q'}} e_ny)_{Q'} 
+ ((1-e_n)x, \frac{P}{P+\overline P} (1-e_n)y)_P\\
&= (e_nx, e_n \frac{Q'}{Q' + \overline{Q'}} e_ny)_{Q'} 
+ ((1-e_n)x, (1-e_n) \frac{P}{P + \overline P} (1-e_n)y)_P, 
\end{align*}
which is compared with the expression for 
$(Q_n' + \overline{Q_n'})(x,y)$ to get 
\[
\frac{Q_n'}{Q_n' + \overline{Q_n'}} 
= \frac{Q'}{Q' + \overline{Q'}} e_n 
+ \frac{P}{P + \overline P} (1-e_n) 
= 
\begin{pmatrix}
1/2 & 0\\
0 & F_n
\end{pmatrix}
\]
and therefore 
\[
\frac{Q_n}{Q_n + \overline{Q_n}} 
= e^{-g_n(F_0-E_0)h_n} 
\frac{Q_n'}{Q_n' + \overline{Q_n'}} 
e^{g_n(F_0-E_0)h_n} 
= 
\begin{pmatrix}
1/2 & f_n\\
0 & F_n
\end{pmatrix}. 
\]
Thus, for $x \in h_nV \subset V$, we have $F_{0,n}x = F_0x$. 
Since 
\[
F_0 e_n = 
\begin{pmatrix}
g_n & 2f({\overline F} - F)h_n\\
0 & 0
\end{pmatrix}
= e_n F_0 e_n, 
\]
it follows that 
\begin{align*}
\varphi_{Q_n,\omega}(e_\omega^{ix}) 
&= e^{i\omega(F_{0,n}x)} 
e^{-Q_n((1-F_{0,n})x)/2}
= e^{i\omega(F_0x)} 
e^{-Q_n((1-F_0)x)/2}\\
&= e^{i\omega(F_0x)} 
e^{-Q((1-F_0)x)/2}
= \varphi_{Q,\omega}(e_\omega^{ix})
\end{align*}
for $x \in h_nV$.   
\end{proof}

\begin{Lemma}
Let $E_0$ (resp.~$F_0$) be the orthogonal projection 
to the subspace $V_0^\C$ with respect to 
$P+\overline P$ (resp.~$Q+\overline Q$) and let 
$D\omega: V/V_0 \to \R$ be defined by 
$\langle D\omega,\dot x\rangle = \omega((E_0-F_0)x)$. 
Set $G = \dot P + \overline{\dot P} 
+ \dot Q + \overline{\dot Q}$
with $G^{-1}$ the inverse form on 
the dual vector space $(V^\C/V_0^\C)^*$. 
Then we have 
\[
(\varphi_{P,\omega}^{1/2}|\varphi_{Q,\omega}^{1/2}) 
= 
\begin{cases}
(\varphi_{\dot P}^{1/2}|\varphi_{\dot Q}^{1/2}) 
e^{-G^{-1}(D\omega)}
&\text{if $D\omega$ is bounded,}\\
0 &\text{otherwise}
\end{cases}
\]
and 
\[
(\varphi_{\dot P}^{1/2}|\varphi_{\dot Q}^{1/2}) 
= \det
\left(
\frac{
\left(
\frac{{\dot Q} + \overline{\dot Q}}
{{\dot P} + \overline{\dot P}}
\right)^{1/2}
+ 
\left(
\frac{{\dot P} + \overline{\dot P}}
{{\dot Q} + \overline{\dot Q}}
\right)^{1/2}
}
{2}
\right)^{-1/2}.
\]
\end{Lemma}

\begin{proof}
We use the notation $G_n$ to stand for 
the positive form ${\dot P} + \overline{\dot P} 
+ {\dot Q}_n + \overline{{\dot Q}_n}$ 
on $(V/V_0)^\C$. 
Let $F_{0,n}$ be the spectral projection 
of $(Q_n+\overline{Q_n})\backslash Q_n$ of an eigenvalue 
$1/2$ and set 
$\langle D_n\omega,\dot x\rangle 
= \omega((E_0 - F_{0,n})x)$ for $\dot x \in (V/V_0)^\C$. 
(Note that $(E_0 - F_{0,n})x \in V_n\cap V_0$.) 

Since the subalgebras $C_n^* = C^*(V_n/\omega_n,\sigma_n)$ of 
$C^*(V/\omega, \sigma)$ 
($\omega_n = \omega|_{V_n \cap V_0}$) 
meet the approximation condition in \cite[Theorem~4.3]{GMS}, 
we see that
\[
(\varphi_{P,\omega}^{1/2}| \varphi_{Q,\omega}^{1/2}) =
\lim_{n \to \infty} 
((\varphi_{P,\omega}|_{C^*(V_n/\omega_n,\sigma_n)})^{1/2} | 
(\varphi_{Q,\omega}|_{C^*(V_n/\omega_n,\sigma_n)})^{1/2}), 
\]
which, in turn, is equal to 
\[
\lim_{n \to \infty} 
((\varphi_{P,\omega}|_{C^*(V_n/\omega_n,\sigma_n)})^{1/2} | 
(\varphi_{Q_n,\omega}|_{C^*(V_n/\omega_n,\sigma_n)})^{1/2}) 
\]
by the above lemma. 

As $P$ and $Q_n$ split according to 
the decomposition $V = V_n \oplus W_n$, we obtain 
\begin{align*}
(\varphi_{P,\omega}^{1/2} | \varphi_{Q_n,\omega}^{1/2})
&= 
((\varphi_{P,\omega}|_{C^*(V_n/\omega_n,\sigma_n)})^{1/2} | 
(\varphi_{Q_n,\omega}|_{C^*(V_n/\omega_n,\sigma_n)})^{1/2})\\ 
&= (\varphi_{\dot P}^{1/2}| \varphi_{{\dot Q}_n}^{1/2}) 
\exp(-G_n^{-1}({\dot D}_n\omega)) 
\end{align*}
and 
\[
(\varphi_{\dot P}^{1/2}| \varphi_{{\dot Q}_n}^{1/2}) 
= \det\left(
\frac{C_n^{1/2} + C_n^{-1/2}}{2} 
\right)^{-1/2}
\quad
\text{with}
\quad 
C_n = \frac{{\dot Q}_n + \overline{{\dot Q}_n}}
{{\dot P} + \overline{\dot P}}.
\]

Likewise, by extracting quotient parts, we have 
\[
(\varphi_{\dot P}^{1/2} | \varphi_{\dot Q}^{1/2}) 
= \lim_{n \to \infty} 
(\varphi_{\dot P}^{1/2} | \varphi_{{\dot Q}_n}^{1/2}). 
\]

Since ${\dot Q}_n \to {\dot Q}$ in the Hilbert-Schmidt 
topology as a part of the convergence 
$Q_n \to Q$, we see 
$C_n \to C = ({\dot P} + \overline{\dot P})\backslash 
({\dot Q} + \overline{\dot Q})$ 
in the Hilbert-Schmidt topology; the determinant formula 
for $(\varphi_{\dot P}^{1/2}|\varphi_{\dot Q}^{1/2})$ 
is proved. 

Since $G_n \to G$ in the Hilbert-Schmidt topology 
and $G$ is invertible with $G^{-1}$ bounded, 
$G_n^{-1} \to G^{-1}$ in the Hilbert-Schmidt topology 
as well. From the expression of 
$(Q_n + \overline{Q_n})\backslash Q_n$ in the previous lemma,
we see
\[
E_0 - F_{0,n} 
= 
\begin{pmatrix}
0 & -2g_nfh_n(\overline{F_n} - F_n)\\
0 & 0
\end{pmatrix}
= 
\begin{pmatrix}
0 & 2f(F - \overline{F})h_n\\
0 & 0
\end{pmatrix}
\]
Since the domain of $\omega$ is chosen so that 
it includes $(E_0-F_0)V$, 
the boundednes of 
$D\omega: \dot x \mapsto \langle \omega, (E_0-F_0)x\rangle$ 
is equivalent to $\omega(E_0-F_0) \in (V/V_0)^*$ 
with $D\omega = \omega(E_0-F_0)$
($(V/V_0)^*$ being furnished with 
a hilbertian topology as the topological dual of 
the hilbertian space $V/V_0$) and hence
$D_n\omega \to D\omega$ in $(V/V_0)^*$. 
Consequently we have 
\[
\lim_{n \to \infty} G_n^{-1}(D_n\omega) 
= G^{-1}(D\omega)
\]
if $D\omega$ is bounded. 

Contrarily assume that $D\omega$ is not bounded.
Since $G_n^{-1} \to G^{-1}$ in the norm topology, 
we can find $\epsilon>0$ so that 
$G_n^{-1} \geq \epsilon({\dot P} + \overline{\dot P})$ for 
$n \geq 1$. 
Then 
\[
\liminf_{n \to \infty} G_n^{-1}(\omega(E_0-F_0)h_n) 
\geq \lim_{n \to \infty} \epsilon 
\| \omega(E_0-F_0)h_n\|_{\dot P}^2 
= +\infty.
\]
\end{proof}

\section{Transition Amplitude Formula}

Our main goal here is a formula  
for the transition amplitude between square roots of 
quasifree states.

Recall that 
two positive forms $A$ and $B$ on a complex vector space $K$ 
is said to be equivalent if we can find 
a positive number $M>0$ 
such that $A(x,x) \leq M B(x,x)$ and 
$B(x,x) \leq MA(x,x)$ for any $x \in K$. 
Equivalent positive forms $A$, $B$ are said to be 
\textbf{HS-equivalent} if ${\mathbf A} - {\mathbf B}$ 
is in the Hilbert-Schmidt class, 
where $\mathbf A$ and 
$\mathbf B$ are operators representing $A$ and $B$ 
on the completion of $K/\ker A = K/\ker B$ 
relative to a positive form 
equivalent to both of $A$ and $B$. 
Note here that the condition is independent of 
the choice of a reference inner product. 

Two polarizations $S$ and $T$ of a presymplectic 
vector space are said to be 
\textbf{equivalent} if positive forms 
$(S^{1/2} + {\overline S}^{1/2})^2$ 
and $(T^{1/2} + {\overline T}^{1/2})^2$ are HS-equivalent. 

\begin{Remark}[{\cite[Proposition~6.6]{AY}}]
The above equivalence on polarizations is equivalent to 
requiring that 
(i) $S + \overline{S}$ and $T + \overline{T}$ are 
equivalent as positive forms and 
(ii) ${\mathbf S}^{1/2} - {\mathbf T}^{1/2}$ 
is in the Hilbert-Schmidt class, 
where $\mathbf S$ and 
$\mathbf T$ are operators representing $S$ and $T$ 
on the completion-after-quotient of $V^\C$ relative 
to a positive form equivalent to $S + \overline{S}$. 
\end{Remark}

\begin{Theorem}\label{infinite}
Let $S$, $T$ be polarizations of 
a presymplectic vector space $(V,\sigma)$ 
with the associated quasifree states denoted by 
$\varphi_S$, $\varphi_T$ 
and define positive forms by 
$2A = (S^{1/2} + {\overline S}^{1/2})^2$, 
$2B = (T^{1/2} + {\overline T}^{1/2})^2$. 
Then we have 
\[
(\varphi_S^{1/2}|\varphi_T^{1/2}) 
= (\varphi_{A/2}^{1/2}|\varphi_{B/2}^{1/2}).
\]
Here the right hand side concerns states 
on the trivial presymplectic vector space 
$(V,0)$, i.e., the gaussian states with covariance forms 
given by $A/2$ and $B/2$. 
\end{Theorem}

\begin{Remark}
The correspondance 
$S \mapsto (S^{1/2} + {\overline S}^{1/2})^2$ is one-to-one. 
\end{Remark}

From the results obtained so far 
(Corollary~4.2, Corollary~4.4 and Lemma~4.6), 
we know that both of the transition amplitudes in question 
are zero if $S$ and $T$ are not equivalent 
as polarizations of $(V,\sigma)$. 
So the equivalence of $S$ and $T$ is assumed 
in the remaining of this section. 
By Proposition~2.8, we may further assume that $V$ is 
non-degenerate and complete 
relative to the inner products $S+\overline S$ and 
$T+\overline T$. 

Recall that the rotated quadrature $P_{\pi/4}$ of $S$ 
is of the form 
\[
P_{\pi/4} = 
\frac{1}{2} 
\begin{pmatrix}
(S^{1/2} + {\overline S}^{1/2})^2 & 
\overline S - S\\
\overline S - S & (S^{1/2} - {\overline S}^{1/2})^2
\end{pmatrix},
\]
where $(S^{1/2} - {\overline S}^{1/2})^2$ is reagrded 
as a positive form on $(V/V_0)^\C$ ($V_0 = \ker \sigma$). 
To avoid confusion, we write $\dot V = V/V_0$ and 
let $\dot S$ be the induced polarization on $\dot V$. 
Then the quotinet form of 
$(S^{1/2} - {\overline S}^{1/2})^2$ on ${\dot V}^\C$ 
admits a continuous extension 
$({\dot S}^{1/2} - {\overline{\dot S}}^{1/2})^2$ 
to the hilbertian completion $W^\C$ of 
${\dot V}^\C$, 
which is the precise meaning of the $(2,2)$-component 
in the above matrix expression of $P_{\pi/4}$. 
Recall also that 
\[
P_{\pi/4} + \overline{P_{\pi/4}} 
= \text{diag}
\left( 
(S^{1/2} + {\overline S}^{1/2})^2, 
({\dot S}^{1/2} - {\overline{\dot S}}^{1/2})^2
\right).
\]
Now the rotated version of $(P+\overline P)\backslash P$ 
is given by 
\begin{align*}
\frac{P_{\pi/4}}{P_{\pi/4}+\overline{P_{\pi/4}}} 
&= \left(
\frac{P_{\pi/4} + \overline{P_{\pi/4}}}{(\ |\ )}
\right)^{-1}
\frac{P_{\pi/4}}{(\ |\ )}\\
&= 
\begin{pmatrix}
(\bS^{1/2} + {\overline\bS}^{1/2})^2 & 0\\
0 & ({\dot \bS}^{1/2} - {\overline{\dot \bS}}^{1/2})^2
\end{pmatrix}^{-1}\\
&\qquad \times \frac{1}{2}
\begin{pmatrix}
(\bS^{1/2} + {\overline\bS}^{1/2})^2 & 
{\overline\bS} - \bS\\
{\overline\bS} - \bS 
& ({\dot \bS}^{1/2} - {\overline{\dot \bS}}^{1/2})^2
\end{pmatrix}\\
&= \frac{1}{2}
\begin{pmatrix}
1_V & 
\frac{{\overline S}^{1/2} - S^{1/2}}
{S^{1/2} + {\overline S}^{1/2}}\\
\frac{S^{1/2} + {\overline S}^{1/2}}
{{\overline S}^{1/2} - S^{1/2}} & 1_W
\end{pmatrix} 
= \frac{1}{2}
\begin{pmatrix}
1_V & -U_S\\
-U_S^* & 1_W
\end{pmatrix}
\end{align*}
Here $(\ |\ )$ denotes the inner product 
$(\ ,\ )_S \oplus (\ ,\ )_{\dot S}$ on 
$V^\C \oplus W^\C$ and the off-diagonal operators 
in the last line are considered to be 
\[
-U_S = \frac{{\overline S}^{1/2} - S^{1/2}}
{S^{1/2} + {\overline S}^{1/2}}: 
W^\C \to V^\C, 
\qquad 
-U_S^* = \frac{S^{1/2} + {\overline S}^{1/2}}
{{\overline S}^{1/2} - S^{1/2}}: 
V^\C \to W^\C.
\]
(Strictly speaking, these involve unbounded operators 
as well as an ill-defined inner product 
$(\ ,\ )_{\dot S}$ on $W^\C$ and therefore they should 
be considered as computations on spectral subspaces of 
$\bS$.)

\begin{Lemma}
Let $P$ and $Q$ be the quadratures of equivalent 
polarizations $S$ and $T$. Then operators 
\[
\frac{P}{P+\overline P} - \frac{Q}{Q+\overline Q}, 
\quad 
\frac{Q+\overline Q}{P+\overline P} - 1
\]
are in the Hilbert-Schmidt class as well.  
\end{Lemma}

\begin{proof}
First $P+\overline P$ and $Q+\overline Q$ are HS-equivalent 
because 
$(S^{1/2} - {\overline S}^{1/2})^2 
- (T^{1/2} - {\overline T}^{1/2})^2$ is in the Hilbert-Schmidt 
class as a quotient of $2(A-B)$. 
Then 
\[
\frac{P}{P+\overline P} - \frac{Q}{Q+\overline Q} 
\sim \frac{P-Q}{P+\overline P} 
= \frac{1}{2} 
\frac{(P+\overline P) - (Q+\overline Q)}
{P+\overline P}
\]
is in the Hilbert-Schimdt class.  
\end{proof}

Let $e_0$ be the $(S + \overline S)$-orthogonal 
projection to the subspace $V_0^\C \subset V^\C$, 
which is the kernel projection of $U_S^*$, i.e, 
$1-e_0 = U_SU_S^*$, and set 
\[
E_0 = 
\begin{pmatrix}
e_0 & 0\\
0 & 0
\end{pmatrix}, 
\qquad 
E_1 = \frac{1}{2}  
\begin{pmatrix}
1_V-e_0 & -U_S\\
-U_S^* & 1_W
\end{pmatrix}. 
\]
The spectral decomposition of 
$(P_{\pi/4} + \overline{P_{\pi/4}})\backslash P_{\pi/4}$ 
is then given by 
\[
\frac{P_{\pi/4}}{P_{\pi/4} + \overline{P_{\pi/4}}} 
= \frac{1}{2} E_0 + E_1. 
\]
The same procedure is applied to $Q$ and $T$ 
to get the spectral decomposition 
$(Q_{\pi/4} + \overline{Q_{\pi/4}})\backslash Q_{\pi/4} 
= F_0/2 + F_1$ with 
\[
F_0 = 
\begin{pmatrix}
f_0 & 0\\
0 & 0
\end{pmatrix}, 
\qquad 
F_1 = \frac{1}{2}  
\begin{pmatrix}
1_V-f_0 & -U_T\\
-U_T^* & 1_W
\end{pmatrix}, 
\]
where $f_0$ denotes the $(T+\overline T)$-orthogonal 
projection to the subspace $V_0^\C \subset V^\C$. 

As in the analysis in \S 5, set $F = (1-e_0)F_1(1-e_0)$. 
Then 
\[
F = 
\begin{pmatrix}
(1_V-e_0)(1_V-f_0)(1_V-e_0) 
& -(1_V-e_0) U_T\\
-U_T^* (1_V - e_0) & 1_W
\end{pmatrix}. 
\]

\begin{Lemma}
We have 
\[
(\varphi_{\dot P}^{1/2}|\varphi_{\dot Q}^{1/2}) 
= (\varphi_{{\dot A}/2}^{1/2}|\varphi_{{\dot B}/2}^{1/2})^2.
\]
\end{Lemma}

\begin{proof}
Instead of passing to the quotient, 
we assume that $V_0 = \{0\}$ to avoid many dots. 
Then in the rotated form 
(the suffix $\pi/4$ being omitted also) 
\begin{align*}
P + \overline P 
&= 
\begin{pmatrix}
(S^{1/2} + {\overline S}^{1/2})^2 & 0\\
0 & (S^{1/2} - {\overline S}^{1/2})^2
\end{pmatrix},\\
Q + \overline Q 
&= 
\begin{pmatrix}
(T^{1/2} + {\overline T}^{1/2})^2 & 0\\
0 & (T^{1/2} - {\overline T}^{1/2})^2
\end{pmatrix}. 
\end{align*}
Thus 
\[
\det\left(
\frac{
\left(
\frac{Q + \overline Q}{P + \overline P}
\right)^{1/2}
+ 
\left(
\frac{P + \overline P}{Q + \overline Q}
\right)^{1/2}}
{2}
\right)
\]
is equal to 
\[
\det\left(
\frac{
(A\backslash B)^{1/2} + (B\backslash A)^{1/2}
}
{2}
\right)
\det\left( 
\frac{C^{1/2} + C^{-1/2}}{2}
\right)
\ \text{with}\  
C = \frac{(S^{1/2} - {\overline S}^{1/2})^2}
{(T^{1/2} - {\overline T}^{1/2})^2} 
\]
and the problem is reduced to showing that 
\[
\det\left(
\frac{
(A\backslash B)^{1/2} + (B\backslash A)^{1/2}
}
{2}
\right)
= \det\left( 
\frac{C^{1/2} + C^{-1/2}}{2}
\right). 
\]
Since both of these determinants are positive, 
it is further reduced to the equality 
\[
\det\left( 
\frac{A\backslash B + B\backslash A + 2}{4}
\right) 
= 
\det\left( 
\frac{C + C^{-1} + 2}{4}
\right).
\]
Now we use the relation
$U_T^* (A \backslash B) U_T = C$ to have 
\[
U_T^*\left( 
\frac{B}{A} + \frac{A}{B} + 2
\right) U_T 
= C + C^{-1} + 2
\]
and we are done by the similarity invariance of determinants. 
\end{proof}

Choose a Hilbert-Schmidt map 
$\Theta: L \to V_0 = \ker\sigma$ 
so that it has a dense range 
and satisfies the condition in \S 5.3 
($\ker \widehat\sigma$ being identified with $V_0$ by 
the canonical isomorphism). 
Recall that 
\[
\frac{1}{4}(E_0 - F_0) = 
\frac{P}{P+\overline P} - 
\left( 
\frac{P}{P+\overline P}
\right)^2 
- \frac{Q}{Q+\overline Q} + 
\left( 
\frac{Q}{Q+\overline Q}
\right)^2 
\]
is in the Hilbert-Schmidt class by Lemma~6.2. 

\begin{Lemma}
For a linear functional $\omega: \Theta L \to \R$, 
the functional $D\omega: {\dot V}\oplus W \to \R$ in Lemma~5.11 
is bounded if and only if 
so is the functional 
$\Delta\omega: {\dot V} \to \R$ in Proposition~3.8, 
and we have 
\[
G^{-1}(D\omega) = ({\dot A} + {\dot B})^{-1}(\Delta\omega). 
\]
\end{Lemma}

\begin{proof}
Recall that the rotated $G_{\pi/4}$ 
is a positive form on ${\dot V}^\C \oplus W^\C$ defined by 
\[
G_{\pi/4}({\dot a} \oplus b, {\dot x} \oplus y) 
= 2({\dot A}+{\dot B})({\dot a},{\dot x}) 
+ (S^{1/2} - {\overline S}^{1/2})^2(b,y)
+ (T^{1/2} - {\overline T}^{1/2})^2(b,y), 
\]
where ${\dot a}, {\dot x} \in {\dot V}^\C$ and 
$b, y \in W$. 
Since an isometric isomorphism 
$v_0: V_0 \to \ker (\sigma \oplus - \sigma)_{\pi/4}$ 
is given by $v_0(x_0) = x_0/\sqrt{2} \oplus 0 \in 
V\oplus W$ and 
$(E_0 - F_0)(\dot x \oplus y) = (e_0-f_0)x \oplus 0$, 
$D\omega$ is equal to the composition 
\[
\begin{CD}
{\dot V} \oplus W @>>> V_0 \oplus W @>>> V_0 @>>> \R\\
@.  @. @. @.\\
{\dot x} \oplus y @>>> (e_0-f_0)x \oplus 0 
@>>> \sqrt{2}(e_0-f_0)x @>>> 
\sqrt{2} \langle \omega, (e_0-f_0)x\rangle  
\end{CD}, 
\]
which should be compared with 
$\langle \Delta\omega, {\dot x}\rangle = \langle \omega, 
(e_0-f_0)x\rangle$. 

To evaluate these functionals by inverse forms, 
we introduce the representing vectors 
${\dot \alpha} \in {\dot V}$ and 
${\dot a} \oplus b \in {\dot V} \oplus W$ by the relation 
\[
\langle \Delta\omega, {\dot x}\rangle 
= ({\dot A} + {\dot B})({\dot \alpha},{\dot x}), 
\langle D\omega, {\dot x} \oplus y\rangle 
= G_{\pi/4}({\dot a}\oplus b,{\dot x} \oplus y). 
\]
The equality 
$\sqrt{2} \langle \Delta\omega,\dot x\rangle 
= \langle D\omega, \dot x \oplus y\rangle$ 
then forces us to have $\sqrt{2}{\dot a} = {\dot\alpha}$,  
$b = 0$ and therefore 
\[
G^{-1}(D\omega) = G({\dot a}\oplus 0) 
= 2({\dot A} + {\dot B})({\dot a}) 
= ({\dot A} + {\dot B})({\dot \alpha}) 
= ({\dot A} + {\dot B})^{-1}(\Delta\omega).
\]
\end{proof}

\textbf{Proof of Theorem:} 
Combining lemmas 5.10, 6.3, 6.4 and 
\cite[Proposition~5.2]{GMS}, 
we have 
\[
(\varphi_{S,\omega}^{1/2}|\varphi_{T,\omega}^{1/2}) 
= (\varphi_{\dot P}^{1/2}|\varphi_{\dot Q}^{1/2})^{1/2} 
e^{-G^{-1}(D\omega)/2}
= (\varphi_{\dot A /2}^{1/2}|\varphi_{\dot B /2}^{1/2})
e^{-({\dot A} + {\dot B})^{-1}(\Delta\omega)}.
\]
Since $A$ and $B$ are HS-equivalent, the spectral 
decomposition of $\frac{B}{A} - 1$ enables us to 
apply Proposition~3.11 to get 
\[
(\varphi_{\dot A /2}^{1/2}|\varphi_{\dot B /2}^{1/2})
e^{-({\dot A} + {\dot B})^{-1}(\Delta\omega)}
= (\varphi_{A/2,\omega}^{1/2}|\varphi_{B/2,\omega}^{1/2}).
\]
Finally we use Corollary~2.11 twice in the following 
to get the formula: 
\begin{align*}
(\varphi_S^{1/2}|\varphi_T^{1/2}) 
&= \int_\Omega \sqrt{\nu_{S_0}\nu_{T_0}}(d\omega) 
(\varphi_{S,\omega}^{1/2}|\varphi_{T,\omega}^{1/2}) 
= \int_\Omega \sqrt{\nu_{S_0}\nu_{T_0}}(d\omega) 
(\varphi_{P,\omega}^{1/2}|\varphi_{Q,\omega}^{1/2})^{1/2}\\ 
&= \int_\Omega \sqrt{\nu_{S_0}\nu_{T_0}}(d\omega) 
(\varphi_{A/2,\omega}^{1/2}|\varphi_{B/2,\omega}^{1/2}) 
= (\varphi_{A/2}^{1/2}|\varphi_{B/2}^{1/2}).
\end{align*}

\appendix

\section{Domination and Ratio Operators}

Let $K$ be a complex vector space and $A$, $B$ be 
sesquilinear forms on $K$. We set 
$A^*(x,y) = \overline{A(y,x)}$ for $x, y \in K$. 
Assume that $B$ is positive, i.e., $B(x,x) \geq 0$ 
for any $x \in K$. A form $A$ is \textbf{dominated} 
by $B$ if there is $\lambda > 0$ such that 
\[
|A(x,y)| \leq \lambda \sqrt{B(x,x) B(y,y)}, 
\quad \forall x, y \in K.
\]
Let $K_B$ be the Hilbert space associated to 
a positive $B$ with the inner product denoted by 
$(\ ,\ )_B$: $K_B$ is the completion of the 
quotient space $K/\ker B$. 
Then we can find a bounded operator $\A: K_B \to K_B$ 
so that 
$A(x,y) = ([x]_B| \A [y]_B)_B$. 
The operator $\A$ is referred to as a ratio operator 
and denoted by 
\[
B \backslash A = \frac{A}{B} = A/B. 
\]
We have $(B\backslash A)^* = B\backslash A^*$, 
where the adjoint operation is taken relative to 
the inner product $(\ |\ )_B$ in $K_B$. 

Assume that $A$ is positive as well. 
Then the condition of domination is equivalent to 
$A(x,x) \leq \lambda B(x,x)$ for all $x \in K$ 
(denoted by $A \leq \lambda B$) by Schwarz' inequality. 
Two positive forms $A, B$ are said to be 
\textbf{equivalent} if each of $A$ and $B$ is dominated by 
the other.
Note that, if this is the case, $K_A = K_B$ and 
the ratio operator $A\backslash B$ is bounded 
with the bounded inverse $B\backslash A$. 
Given a bounded linear operator $O$ on $K_A = K_B$, 
let $O^A$ (resp.~$O^B$) be the adjoint of $O$ 
with respect to the inner product 
$(\ |\ )_A$ (resp.~$(\ |\ )_B$). 
Then we have 
\[
O^B = \frac{A}{B} O^A \frac{B}{A}.
\]
Particularly $O = A\backslash B$ is self-adjoint relative 
to $(\ |\ )_B$ and 
\[
\left\| 
\frac{B}{A} 
\right\|_A 
= \sup\left\{ 
\frac{\| x\|_B}{\| x\|_A}; 
0 \not= x \in K_A = K_B
\right\}
= \left\| 
\frac{A}{B} 
\right\|_B^{-1}.
\]

\begin{Remark}
The operator $A\backslash B$ is positive relative to 
$(\ |\ )_B$ because $(A \backslash B)^{1/2}$ is 
$B$-hermitian. 
\end{Remark}

Let $R = (A \backslash B)^{1/2}$ and regard it 
as a unitary map 
$(K_B,(\ |\ )_B) \to (K_A,(\ |\ )_A)$. 
Then, for a bounded operator $O$ on $K_A = K_B$, 
the commutativity of the diagram 
\[
\begin{CD}
K_B @>{R}>> K_A\\
@V{O}VV @VV{ROR^{-1}}V\\
K_B @>>{R}> K_A
\end{CD}
\]
shows that $\| O\|_B = \| ROR^{-1}\|_A$ and 
\[
\sum_j \| O\eta_j\|_B^2 = 
\sum_j \| ROR^{-1} R\eta_j \|_A^2 
= \sum_j \| ROR^{-1}\xi_j\|_A^2
\]
with $\{ \xi_j \}$ and $\{ \eta_j \}$ orthonormal bases 
for $(\ |\ )_A$ and $(\ |\ )_B$ respectively. 

\begin{Lemma}
Let $C$ be a sesquilinear form. Then 
\[
(A\backslash B)(B\backslash C) = A\backslash C.
\]
\end{Lemma}

\begin{proof}
Just compute as 
\[
(x|\frac{B}{A}\frac{C}{B}y)_A 
= (x|\frac{C}{B}y)_B 
= C(x,y) = (x|\frac{C}{A}y)_A
\quad
\text{for $x, y \in K_A = K_B$.} 
\]
\end{proof}

\begin{Remark}
Notice that $\frac{C}{B}\frac{B}{A} \not= \frac{C}{A}$ 
generally: The backslash notation 
is therefore safer in applying cancellation. 
\end{Remark}

Given equivalent positive forms $A, B$, 
we introduce the projective 
distance $\delta(A,B)$ between them by 
\[
\delta(A,B) = \inf 
\{ \log(\lambda \mu); A \leq \lambda B, 
B \leq \mu A \}.
\]
The projective distance is in fact a distance on 
the set of rays and we have $\delta(A,B) \geq 0$ and 
$\delta(A,B) = 0$ if and only if $A$ and $B$ are 
proportional.

\begin{Definition}
Let $Q$ and $Q'$ be positive quadratic forms on 
a finite-dimensional real vector space 
and assume that $Q'$ is dominated by $Q$. 
Then the \textbf{relative determinant} is defined by 
\[
\det\left( 
\frac{Q'}{Q} 
\right) = 
\frac{\det_{1\leq j,k \leq n}(Q'(v_j,v_k))}
{\det_{1\leq j,k \leq n}(Q(v_j,v_k))}. 
\]
Here $\{ v_j \} \subset V$ is any representative family of 
basis in $V'$. 
\end{Definition}

\section{Gaussian Measures}
We shall review here relevant results on gaussian measures, 
which can be found in standard textbooks such as 
\cite{Si1, J}. 

Let $\R^\infty$ be the set of sequences of real numbers 
and $\sB(\R^\infty)$ be the product Borel structure. 
Given a sequence $\alpha = (\alpha_n)_{n \geq 1}$ of positive 
numbers, let $\nu_\alpha$ be the infinite product measure 
of gaussian measures of variance $\alpha_j$ ($j \geq 1$): 
If we denote by $X_j: \R^\infty \to \R$ the projection 
random variable to $j$-th component, then 
\[
\int_{\R^\infty} \nu_\alpha(dx)\, 
e^{i\sum_{j=1}^n t_jX_j(x)} 
= e^{-\sum_{j=1}^n\alpha_j t_j^2/2}.
\]

\begin{Proposition}
Let $\alpha, \beta \in \R_+^\infty$ and set 
\[
\R_\beta^\infty = 
\{ x = (x_j) \in \R^\infty; \sum_{j=1}^\infty \beta_j x_j^2 < +\infty 
\}.
\]
Then 
\[
\nu_\alpha(\R_\beta^\infty) = 
\begin{cases}
1 &\text{if $\sum_j \alpha_j\beta_j < +\infty$,}\\
0 &\text{otherwise.}
\end{cases}
\]
In other words, 
$\sum_j \beta_j x_j^2 < +\infty$ for $\nu_\alpha$-a.e.~ 
$x \in \R^\infty$ if $\sum_j \alpha_j\beta_j < +\infty$, 
and 
$\sum_j \beta_j x_j^2 = +\infty$ for $\nu_\alpha$-a.e.~ 
$x \in \R^\infty$ if $\sum_j \alpha_j\beta_j = +\infty$.  
\end{Proposition}

Given an admissible positive form $S$ 
on a separable real hilbertian space $V$, 
we can realize 
the gaussian random process indexed by $(V,S)$ 
in terms of the topological dual 
$(\Theta V)^*$ of $\Theta V$. Here 
$\Theta: V \to V$ is any invertible operator in 
the Hilbert-Schmidt class and 
$\Theta V$ is a hilbertian space so that 
$V \ni x \mapsto \Theta x \in \Theta V$ is a topological 
isomorphism. Note also that the topological dual 
$(\Theta V)^*$ is identified with $V^*\Theta^{-1}$: 
any $f \in (\Theta V)^*$ is of the form 
$f(\Theta x) = g(x)$ ($x \in V$) with $g \in V^*$.  

Let $\sB$ be the Boolean algebra of Borel sets 
in $V^*\Theta^{-1}$. 

\begin{Proposition}
We can find a probability measure $\nu_S$ on 
$((\Theta V)^*,\sB)$ such that 
\[
\int_{(\Theta V)^*} e^{i\langle x,\omega\rangle} \nu_S(d\omega) 
= e^{-S(x)/2}
\]
and 
\[
\int_{(\Theta V)^*} \langle x,\omega\rangle 
\langle y,\omega\rangle \nu_S(d\omega) 
= S(x,y)
\]
for $x, y \in \Theta V$. 

Furthermore, the correspondance 
$\Theta V \ni x \mapsto \langle x, \cdot \rangle \in 
L^2((\Theta V)^*,\nu_S)$ is 
exntended to a gaussian random process 
$V \to L^2((\Theta V)^*,\nu_S)$ indexed by $(V,S)$. 
In other words, $x \mapsto \langle x, \cdot\rangle$ 
is continuous relative to the subtopology of $V$: 
If $\lim x_n = x$ in $V$ with $x_n \in \Theta V$ and $x \in V$, 
then $\{ \langle x_n,\cdot\rangle\}$ is convergent 
in $L^2(\Omega,\nu_S)$. 
\end{Proposition}

\end{document}